\documentclass[english,12pt,onecolumn, draftcls]{IEEEtran}
\usepackage[T1]{fontenc}
\usepackage[latin9]{inputenc}
\usepackage{amsthm}
\usepackage{amsmath}
\usepackage{graphicx}
\usepackage{amssymb}

\makeatletter
\theoremstyle{plain}

\theoremstyle{plain}
\newtheorem{prop}{Proposition}
\theoremstyle{remark}
\newtheorem{rem}{Remark}
\theoremstyle{plain}

\theoremstyle{plain}
\newtheorem{lem}{Lemma}

\ifCLASSINFOpdf
\else
\fi
\usepackage{babel}

\makeatother

\usepackage{babel}

\begin{document}

\title{Source Coding When the Side Information May Be Delayed}

\author{Osvaldo Simeone,~\IEEEmembership{Member,~IEEE,} and~Haim Permuter,~\IEEEmembership{Member,~IEEE}
\thanks{O. Simeone is with the Center for Wireless Communications and Signal
Processing Research (CWCSPR), ECE Department, New Jersey Institute
of Technology (NJIT), Newark, NJ 07102, USA (email: osvaldo.simeone@njit.edu).
H. H. Permuter is with the Department of Electrical and Computer Engineering,
Ben-Gurion University of the Negev, Beer-Sheva 84105, Israel (e-mail:
haimp@bgu.ac.il) %
}%
\thanks{The work of O. Simeone was supported in part by the U.S. National
Science Foundation under Grant No. 0914899. H. H. Permuter was supported
in part by the Marie Curie Reintegration fellowship.

This work was presented in part at IEEE International Symposium on
Information Theory (ISIT), July 2012, Cambridge, MA, USA.%
}}

\maketitle

\begin{abstract}
For memoryless sources, delayed side information at the decoder does
not improve the rate-distortion function. However, this is not the
case for sources with memory, as demonstrated by a number of works
focusing on the special case of (delayed) feedforward. In this paper,
a setting is studied in which the encoder is potentially uncertain
about the delay with which measurements of the side information, which
is available at the encoder, are acquired at the decoder. Assuming
a hidden Markov model for the source sequences, at first, a single-letter
characterization is given for the set-up where the side information
delay is arbitrary and known at the encoder, and the reconstruction
at the destination is required to be asymptotically lossless. Then,
with delay equal to zero or one source symbol, a single-letter characterization
of the rate-distortion region is given for the case where, unbeknownst
to the encoder, the side information may be delayed or not, and additional
information can be received by the decoder when the side information
is not delayed. Finally, examples for binary and Gaussian sources
are provided.
\end{abstract}
\begin{IEEEkeywords} Rate-distortion function, Hidden Markov Model,
Markov Gaussian process, multiplexing, strictly causal side information,
causal conditioning. \end{IEEEkeywords}

\section{Introduction}

Consider a sensor network in which a sensor measures a certain physical
quantity $Y_{i}$ over time $i=1,2,...n.$ The aim of the sensor is
communicating a symbol-by-symbol processed version $X^{n}=(X_{1},...,X_{n})$
of the measured sequence $Y^{n}=(Y_{1},...,Y_{n})$ to a receiver.
As an example, each element $X_{i}$ can be obtained by quantizing
or denoising $Y_{i}$, for $i=1,2,...n.$ To this end, based on the
observation of $X^{n}$ and $Y^{n}$, the sensor communicates a message
$M$ of $nR$ bits to the receiver ($R$ is the message rate in bits
per source symbol). The receiver is endowed with sensing capabilities,
and hence it can measure the physical quantity $Y^{n}$ as well. However,
as the receiver is located further away from the physical source,
such measure may come with some delay, say $n+d$ for some $d\geq0$.
Assuming that at time $n+i$ the decoder must put out an estimate
$Z_{i}$ of the $i$th source symbol $X_{i}$ by design constraints,
it follows that the estimate $Z_{i}$ can be made to be a function
of the message $M$ and of the delayed side information $Y^{i-d}=(Y_{1},...,Y^{i-d})$
(see \cite{pradhan} for an illustration). Following related literature
(e.g., \cite{pradhan directed}), we will refer to $d$ as the delay
for simplicity. Delay $d$ may or may not be known at the sensor.

\begin{figure}[!t]
\centering \includegraphics[width=4.5in]{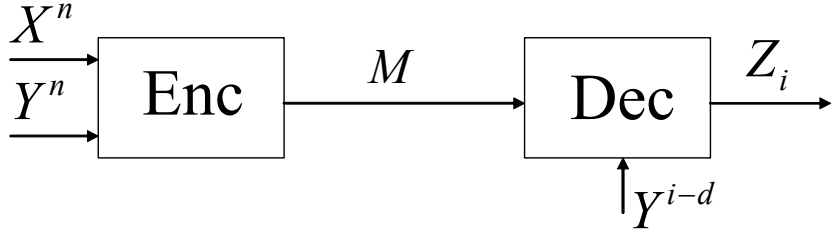} \caption{Source coding with delayed side information at the decoder. The side
information is fully available at the encoder.}

\label{fig0}
\end{figure}

The situation described above can be illustrated schematically as
in Fig. \ref{fig0} for the case in which the delay $d$ is known
at the encoder. In Fig. \ref{fig0}, the encoder (\textquotedbl{}Enc\textquotedbl{})
represents the sensor and the decoder (\textquotedbl{}Dec\textquotedbl{})
the receiver. The decoder at time $i$ (more precisely, $n+i$) has
access to delayed \textit{side information} $Y^{i-d}$ with delay
$d.$ Fig. \ref{fig2} accounts for a setting where the side information
at the decoder, unbeknownst to the encoder, \textit{may }be delayed
by $d$ or not delayed, where the first case is modelled by Decoder
1 and the second by Decoder 2. Note that, in the latter case, the
receiver has available the sequence $Y^{i}=(Y_{1},...,Y_{i})$ at
time $i$. For generality, in the setting in Fig. \ref{fig2}, we
further assume that the encoder is allowed to send additional information
in the form of a message $M_{\Delta}$ of $n\Delta R$ bits when the
side information is not delayed. This can be justified in the sensor
example mentioned above, as a non-delayed side information may entails
that the receiver is closer to the transmitter and is thus able to
decode an additional message of rate $\Delta R$ (bits/source symbol).

\subsection{Preliminary Considerations and Related Work\label{sub:Preliminary-Considerations-and}}

\begin{figure}
\begin{centering}
\centering\includegraphics[width=4.5in]{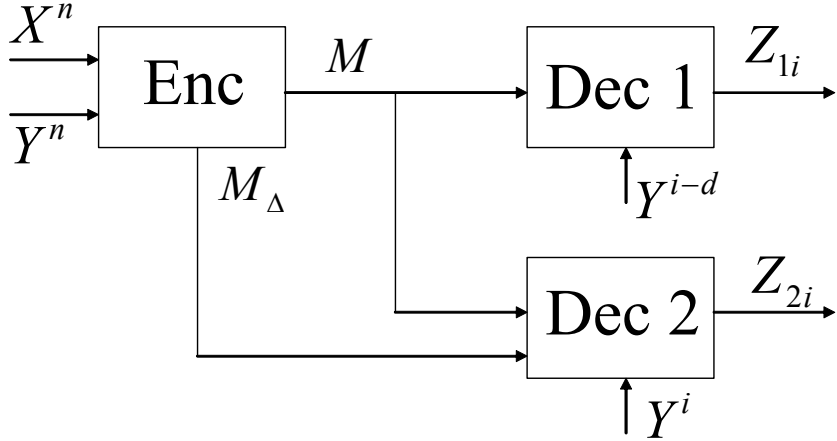}
\par\end{centering}

\caption{Source coding where side information at the decoder may be delayed
and additional information can be delivered when side information
is not delayed. The side information is fully available at the encoder.}

\label{fig2}
\end{figure}

To start, let us first assume that sequences $X^{n}$ and $Y^{n}$
are \emph{memoryless sources }so that the entries ($X_{i},Y_{i}$)
are arbitrarily correlated for a given index $i$ but independent
identically distributed (i.i.d.) for different $i=1,...,n.$ To streamline
the discussion, the following lemma summarizes the optimal trade-off
between rate $R$ and distortion $D$, as measured by a distortion
metric $d(x,z)$, for the point-to-point setting of Fig. \ref{fig0}
with memoryless sources. Similar conclusions apply for the more general
set-up of Fig. \ref{fig2}.
\begin{lem}
\cite{Berger,Gray conditional,Weissman coding} For memoryless source,
and zero delay, i.e., $d=0$, the rate-distortion function for the
point-to-point system in Fig. \ref{fig0} is given by the conditional
rate-distortion function
\begin{equation}
R(D)=\min_{p(z|x,y):\textrm{ }\textrm{\ensuremath{\mathrm{E}}\ensuremath{[d(X,Z)]\leq D}}}I(X;Z|Y).
\end{equation}
 This result remains unchanged even if the decoder has access to non-causal
side information, i.e., if the reconstruction $Z_{i}$ can be based
on the entire sequence $Y^{n}$, rather than only $Y^{i}$. Instead,
for strictly positive delay $d>0$, the rate-distortion function is
the same as if there was no side information, namely $R(D)=\min_{p(z|x):\textrm{ }\textrm{\ensuremath{\mathrm{E}}\ensuremath{[d(X,Z)]\ensuremath{\leq}D}}}I(X;Z)$.%
\footnote{The first part of the Lemma is due to \cite{Berger,Gray conditional},
while the second can be derived as in \cite[Observation 2]{Weissman coding}. %
}
\end{lem}

Similar conclusions can be easily shown to apply also for the more
general model of Fig. \ref{fig2}, as it will be discussed in the
paper (see Sec. \ref{sec:Lossy-Source-Coding}). Specifically, if
$d>0$ and the sources are memoryless, the rate-distortion function
for the system of Fig. 2 with $\Delta R=0$ reduces to the one obtained
by Kaspi in \cite{Kaspi} for a model in which decoder 1 has no side
information, and, for general $\Delta R\geq0$ , the rate-distortion
region coincides with the one obtained in \cite{maor1} for a model
with no side information at decoder 1.

We have seen in Lemma 1 that, for memoryless sources, no advantages
can be accrued by leveraging a (strictly) delayed side information,
i.e., with $d>0.$ However, this conclusion does not generally hold
if the sources have memory. In this context, a number of works have
focused on the scenario of Fig. \ref{fig0} where $X_{i}=Y_{i}$ for
$i=1,...n.$ This entails that the decoder observes sequence $X^{n}$
itself, but with a delay of $d$ symbols. This setting is typically
referred to as \textit{source coding with feedforward}, and was introduced
in \cite{weissman03}. Reference \cite{pradhan} derived the rate-distortion
function for this problem (i.e., Fig. \ref{fig0} with $X_{i}=Y_{i}$)
for ergodic and stationary sources in terms of multi-letter mutual
informations. The result was also extended to arbitrary sources using
information-spectrum methods. Achievability was obtained via the use
of a codebook of codetrees. The function was explicitly evaluated
for some special cases in \cite{Naiss,El Gamal} (see also \cite{pradhan evaluating}),
and \cite{Naiss} proposed an algorithm for its numerical calculation.

The more general case of Fig. \ref{fig0} with $X_{i}\neq Y_{i}$
was studied in \cite{pradhan directed} assuming stationary and ergodic
sources $X^{n}$ and $Y^{n}$. The rate-distortion function was expressed
in terms of multi-letter mutual informations. No specific examples
were provided for which the function is explicitly computable. We
finally remark that for more complex networks than the ones studied
here, strictly delayed side information may be useful even in the
presence of memoryless sources. This was illustrated in \cite{pradhan md 1}
for a multiple description problem with feedforward.

\begin{figure}
\begin{centering}
\includegraphics[width=4.5in]{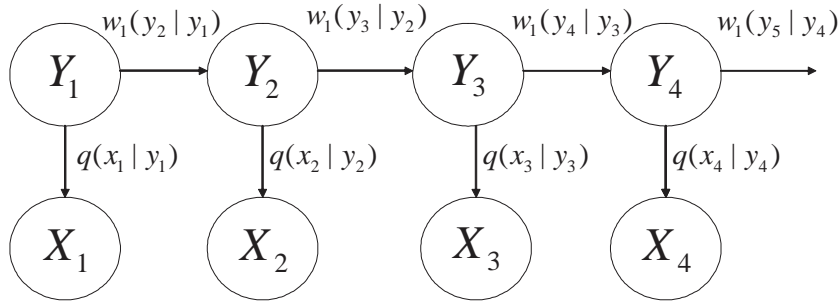}
\par\end{centering}

\caption{A graphical illustration of the assumed hidden Markov model for the
sources.}

\label{figmarkov}
\end{figure}

\subsection{Contributions}

The goal of this work is to characterize the rate-distortion trade-offs
for the setting in Fig. \ref{fig0} and the more general set-up in
Fig. \ref{fig2} for a specific class of sources $X^{n}$ and $Y^{n}$.
Specifically, we assume that $Y^{n}$ is a Markov chain, and $X^{n}$
is such that $X_{i}$ is obtained by passing $Y_{i}$ through a channel
$q(x|y)$ for $i=1,...,n,$ as illustrated in Fig. \ref{figmarkov}.
The process is thus a hidden Markov model. This model complies with
the type of sensor network scenarios described above, where $Y^{n}$
is the physical quantity of interest, modelled as a Markov chain,
and $X^{n}$ is a symbol-by-symbol processed version of $Y^{n}.$
The main contributions and the paper organization are as follows.
After the description of the system model in Sec. \ref{sec:System-Model},
for the source statistics described above,
\begin{itemize}
\item we derive a single-letter characterization of the minimal rate (bits/source
symbol) required for asymptotically lossless compression in the point-to-point
model of Fig. \ref{fig0} for any delay $d\geq0$ (Sec. \ref{sec:Lossless-Source-Coding}).
Achievability is based on a novel scheme that consists of simple multiplexing/demultiplexing
operations along with standard entropy coding techniques;
\item we derive a single-letter characterization of the minimal rate (bits/source
symbol) required for lossy compression for the point-to-point model
of Fig. \ref{fig0} and, more generally, for the model of Fig. \ref{fig2}
in which the side information may be delayed, for delays $d=0$ and
$d=1$ (Sec. \ref{sec:Lossy-Source-Coding});
\item we solve a number of specific examples, namely binary-alphabet sources
with Hamming distortion and Gaussian sources with minimum mean square
error distortion, and present related numerical results (Sec. \ref{sec:Examples}).
\end{itemize}

\section{System Model\label{sec:System-Model}}

We present the system model for the scenario of Fig. \ref{fig2}.
As detailed below, the scenarios of Fig. \ref{fig0} is obtained as
a special case. The system is characterized by a delay $d\geq0$;
finite alphabets $\mathcal{X}$, $\mathcal{Y}$, $\mathcal{Z}_{1}$,
$\mathcal{Z}_{2};$ conditional probabilities $w_{1}(a|b)$, with
$a,b\in\mathcal{Y},$ and $q(x|y),$ with $x\in\mathcal{X}$ and $y\in\mathcal{Y}$
(i.e., we have $\sum_{a\in\mathcal{Y}}w_{1}(a|b)=1$ and $\sum_{a\in\mathcal{X}}q(a|b)=1$
for all $b\in\mathcal{Y}$); and distortion metrics $d_{j}(x,y,z_{j})$:
$\mathcal{X}\times\mathcal{Y}\times\mathcal{Z}_{j}\rightarrow[0,d_{\max}]$,
such that $0\leq\mathrm{d}_{j}(x,y,z_{j})\leq\mathrm{d}_{\max}<\infty$
for all $(x,y,z)\in\mathcal{X}\times\mathcal{Y}\times\mathcal{Z}_{j}$
for $j=1,2$. As explained below, the subscript {}``1'' in $w_{1}(a|b)$
indicates that $w_{1}(a|b)$ denotes one-step transition probabilities.

The random process $Y_{i}\in\mathcal{Y}$, $i\in\{...,-1,0,1,...\}$,
is a stationary and ergodic Markov chain with transition probability
$\Pr[Y_{i}=a|Y_{i-1}=b]=w_{1}(a|b).$ We define the probability $\Pr[Y_{i}=a]\triangleq\pi(a)$
and also the $k$-step transition probability $\Pr[Y_{i}=a_{i}|Y_{i-k}=b]\triangleq w_{k}(a|b),$
which are both independent of $i$ by the stationarity of $Y_{i}$.
These quantities can be calculated using standard Markov chain theory
from the transition matrix associated with $w_{1}(a|b)$ (see, e.g.,
\cite{Gallager}). We also set, for notational convenience, $w_{0}(a|b)=\pi(a)$.
Sequence $Y^{n}=(Y_{1},...,Y_{n})$ is thus distributed as $p(y^{n})=\pi(y_{1}\mbox{) }\prod\nolimits _{i=2}^{n}w_{1}(y_{i}|y^{i-1})$
for any integer $n>0.$

The random process $X_{i}\in\mathcal{X}$, $i\in\{...,-1,0,1,...\}$
is such that vector $X^{n}=(X_{1},...,X_{n})\in\mathcal{X}^{n}$,
for any integer $n>0$, is jointly distributed with $Y^{n}$ so that
\begin{align}
p(x^{n},y^{n}) & =\pi(y_{1}\mbox{)\ensuremath{q(x_{1}|y_{1})}}\prod\limits _{i=2}^{n}p(x_{i},y_{i}|x^{i-1},y^{i-1})\nonumber \\
 & =\pi(y_{1}\mbox{)\ensuremath{q(x_{1}|y_{1})}}\prod\limits _{i=2}^{n}w_{1}(y_{i}|y^{i-1})q(x_{i}|y_{i}).\label{eq:hidden Markov}
\end{align}
 In other words, process $X_{i}\in\mathcal{X}$, $i\in\{...,-1,0,1,...\}$
corresponds to a hidden Markov model with underlying Markov process
given by $Y^{n}.$

We now define encoder and decoders for the setting of Fig. \ref{fig2}.
Specifically, an $(d,n,R,\Delta R,D_{1},D_{2})$ code is defined by:
(\textit{i}) An encoder function
\begin{equation}
\mathrm{f}\text{: }(\mathcal{X}^{n}\times\mathcal{Y}^{n})\rightarrow\lbrack1,2^{nR}]\times\lbrack1,2^{n\Delta R}],\label{eq:encoder}
\end{equation}
 which maps sequences $X^{n}$ and $Y^{n}$ into messages $M\in\lbrack1,2^{nR}]$
and $M_{\Delta}\in\lbrack1,2^{n\Delta R}];$ (\textit{ii}) a sequence
of decoding functions for decoder 1
\begin{equation}
\mathrm{g}_{1i}\text{: }[1,2^{nR}]\times\mathcal{Y}^{i-d}\rightarrow\mathcal{Z}_{1},\label{eq:decoder}
\end{equation}
 for $i\in\lbrack1,n]$, which, at each time $i,$ map message $M,$
or rate $R$ {[}bits/source symbol{]}, and the delayed side information
$Y^{i-d}$ into the estimate $Z_{1i}$; (\textit{iii}) a sequence
of decoding function for decoder 2
\begin{equation}
\mathrm{g}_{2i}\text{: }[1,2^{nR}]\times[1,2^{n\Delta R}]\times\mathcal{Y}^{i}\rightarrow\mathcal{Z}_{2}\label{eq:decoder2}
\end{equation}
 for $i\in\lbrack1,n]$, which, at each time $i,$ map messages $M,$
or rate $R,$ and $M_{\Delta},$ of rate or rate $\Delta R,$ and
the non-delayed side information $Y^{i}$ into the estimate $Z_{2i}$.
In (\ref{eq:encoder})-(\ref{eq:decoder2}), for $a,b$ integer with
$a\leq b$, we have defined $[a,b]$ as the interval $[a,a+1,...,b]$
with $[a,b]=\phi$ if $a>b$.%
\footnote{As it is standard practice, $2^{nR}$ and $2^{n\Delta R}$ are implicitly
considered to be rounded up to the nearest larger integer. %
} Encoding/decoding functions (\ref{eq:encoder})-(\ref{eq:decoder2})
must satisfy the distortion constraints
\begin{equation}
\frac{1}{n}\sum\limits _{i=1}^{n}\mathrm{E}[\mathrm{d}_{j}(X_{i},Y_{i},Z_{ji})]\leq D_{j},\text{ for }j=1,2.\label{dist constraints}
\end{equation}
 Note that these constraints are fairly general in that they allow
to impose not only requirements on the lossy reconstruction of $X_{i}$
or $Y_{i}$ (obtained by setting $\mathrm{d}_{j}(x,y,z_{j})$ independent
of $y$ or $x,$ respectively), but also on some function of both
$X_{i}$ and $Y_{i}$ (by setting $\mathrm{d}_{j}(x,y,z_{j})$ to
be dependent on such function of ($x,y$)).

Given a delay $d\geq0$, for a distortion pair ($D_{1},D_{2}$), we
say that rate pair ($R,\Delta R$) is achievable if, for every $\epsilon>0$
and sufficiently large $n$, there exists a $(d,n,R,\Delta R,D_{1}+\epsilon,D_{2}+\epsilon)$
code. We refer to the closure of the set of all achievable rates for
a given distortion pair ($D_{1},D_{2}$) and delay $d$ as the \emph{rate-distortion
region} $\mathcal{R}_{d}(D_{1},D_{2})$.

From the general description above for the setting of Fig. \ref{fig2},
the special case of Fig. \ref{fig0} is produced by neglecting the
presence of decoder 2, or equivalently by choosing $D_{2}=\mathrm{d}_{\max}$.
In this case, the rate-distortion region $\mathcal{R}_{d}(D_{1},D_{2})$
is fully characterized by a function $R_{d}(D_{1})$ as $\mathcal{R}_{d}(D_{1},\mathrm{d}_{\max})$$=\{(R,\Delta R):\textrm{ \ensuremath{R\geq R_{d}(D_{1}),\mbox{ \ensuremath{\Delta R\geq0\}}}}}$.
Function $R_{d}(D_{1})$ hence characterizes the infimum of rates
$R$ for which the pair $(D_{1},\mathrm{d}_{\max})$ is achievable,
and is referred to as the \emph{rate-distortion function} for the
setting of Fig. \ref{fig0}. For the special case of the model in
Fig. \ref{fig2} in which $\Delta R=0$, we define the rate-distortion
function $R_{d}(D_{1},D_{2})$ in a similar way.

\textit{Notation}: For $a,b$ integer with $a<b$, we define $x_{a}^{b}=(x_{a},...,x_{b})$;
if instead $a<b$ we set $x_{a}^{b}=\emptyset$. We will also write
$x_{1}^{b}$ for $x^{b}$ for simplicity of notation. Given a sequence
$x^{n}=[x_{1},...,x_{n}]$ and a set $\mathcal{I}=\{i_{1},...,i_{|\mathcal{I}|}\}\subseteq[1,n],$
we define sequence $x^{\mathcal{I}}$ as $x^{\mathcal{I}}=[x_{i_{1}},x_{i_{2}},...,x_{i_{|\mathcal{I}|}}]$
where $i_{1}\leq...\leq i_{|\mathcal{I}|}$. Random variables are
denoted with capital letters and corresponding values with lowercase
letters. Given random variables, or more generally vectors, $X$ and
$Y$ we will use the notation $p_{X}(x)$ or $p(x)$ for $\Pr[X=x]$,
and $p_{X|Y}(x|y)$ or $p(x|y)$ for $\Pr[X=x|Y=y]$, where the latter
notations are used when the meaning is clear from the context. Given
set $\mathcal{X}$, we define $\mathcal{X}^{n}$ as the $n$-fold
Cartesian product of $\mathcal{X}$. We denote any function of $\epsilon>0$
that tends to zero as $\epsilon\rightarrow0$ as $\delta(\epsilon)\rightarrow0$.
When referring to $\epsilon-$typical sequences, we refer to the notion
of strong typicality as treated in \cite{El Gamal Kim}.

\section{Point-to-Point Model}

In this section, we study the point-to-point model in Fig. \ref{fig0}.

\subsection{Lossless Compression\label{sec:Lossless-Source-Coding}}

We start by characterizing the rate-distortion function $R_{d}(D_{1})$
for any delay $d\geq0$ under the Hamming distortion metric for $D_{1}=0$.
The Hamming distortion metric is defined as $\mathrm{d}_{1}(x,y,z_{1})=\mathrm{1}(x\neq z_{1})$,
where $\mathrm{1}(a)=1$ if $a$ is true and $\mathrm{1}(a)=0$ otherwise.
This implies that the distortion constraint (\ref{dist constraints})
for $j=1$ becomes
\begin{equation}
\frac{1}{n}\sum\limits _{i=1}^{n}\mathrm{E}[1(X_{i}\neq Z_{1i})]=\frac{1}{n}\sum\limits _{i=1}^{n}\Pr[X_{i}\neq Z_{1i}]=0.\label{dist constraints-1}
\end{equation}
 In other words, from the definition of achievability given above,
we impose that the sequence $X^{n}$ be recovered with vanishingly
small average symbol error probability as $n\rightarrow\infty$. We
refer to this scenario as asymptotically lossless, or lossless for
short.

We have the following characterization of $R_{d}(0)$.
\begin{prop}
\label{prop:For-any-delay 1}For any delay $d\geq0$, the rate-distortion
function for the set-up in Fig. \ref{fig0} under Hamming distortion
at $D_{1}=0$ is given by
\begin{equation}
R_{d}(0)=H(X_{d+1}|X_{2}^{d},Y_{1}),\label{lossless}
\end{equation}
 where the conditional entropy is calculated with respect to the distribution
\begin{align}
p(y_{1},x_{1}) & =\pi(y_{1})q(x_{1}|y_{1})\text{ for }d=0,\label{eq: distributions-a}\\
\text{and }p(y_{1},x_{2},...,x_{d+1}) & =\pi(y_{1})\sum_{\substack{y_{i}\in\mathcal{Y}\\
i\in\lbrack2,d+1]
}
}\prod\limits _{i=2}^{d+1}w_{1}(y_{i}|y_{i-1})q(x_{i}|y_{i})\text{ for }d\geq1.\label{eq:distributions}
\end{align}

\end{prop}
The proof of converse of the proposition above is based on an appropriate
use of the Fano inequality and is reported in Appendix \ref{sub:Proof-of-Converse}.
To prove the direct part of the proposition, we propose a simple achievable
scheme, which, to the best of the authors' knowledge, has not appeared
before, in Sec. \ref{sub:Proof-of-Achievability}.

\begin{rem} Expression (\ref{lossless}) consists of a conditional
entropy of $d+1$ random variables, namely $Y_{1}$,$X_{2}$, ...,
$X_{d+1}$. These variables are distributed as the corresponding entries
in the random vectors $X^{n}$ and $Y^{n}$, as per (\ref{eq: distributions-a})-(\ref{eq:distributions})
(cf. (\ref{eq:hidden Markov})). We have therefore used the same notation
for the involved random variables as in Sec. \ref{sec:System-Model}.
Proposition 1 provides a {}``single-letter'' characterization of
$R_{d}(0)$ for the setting of Fig. \ref{fig0}, since it only involves
a finite number of variables%
\footnote{It might be more accurately referred to as a {}``finite-letter''
characterization.%
}. This contrasts with the general characterization for stationary
ergodic processes of $R_{d}(D)$ given in \cite{pradhan directed},
which is a {}``multi-letter'' expression, whose computation can
generally only attempted numerically using approaches such as the
ones proposed in \cite{Naiss}. Note that a multi-letter expression
is also given in \cite{El Gamal} to characterize $R_{d}(D)$ for
i.i.d. sources with \emph{negative} delays $d<0$. Finally, it should
be emphasized that the simple characterization (\ref{lossless}) for
the scenario of interest here hinges on the assumed statistics of
the sources ($X^{n},Y^{n}$).\end{rem}

\begin{rem}By setting $d=0$ in (\ref{lossless}) we obtain $R_{0}(0)=H(X_{1}|Y_{1})$.
This result generalizes \cite[Remark 3, p. 5227]{El Gamal} from i.i.d.
sources ($X^{n},Y^{n}$) to the hidden Markov model (\ref{eq:hidden Markov})
considered here. Note that, for $d=1$, we instead obtain $R_{1}(0)=H(X_{2}|Y_{1}).$
As another notable special case, if side information is absent, or
equivalently $d\rightarrow\infty$, in accordance to well-known results,
we obtain that $R_{\infty}(0)$ equals the entropy rate (see, e.g.,
\cite{Cover})
\begin{equation}
H(\mathcal{X})\triangleq\lim_{n\rightarrow\infty}H(X_{1},..,X_{n}).
\end{equation}
In fact, we have
\begin{equation}
R_{\infty}(0)=\lim_{d\rightarrow\infty}H(X_{d+1}|X_{2}^{d},Y_{1})=H(\mathcal{X})\label{eq:entropy in}
\end{equation}
 by \cite[Theorem 4.5.1]{Cover}.\end{rem}

\begin{rem}\label{rem:Is-delayed-side}Is \emph{delayed} side information
useful (when known also at the encoder)? That this is generally the
case follows from the inequality
\begin{equation}
R_{d}(0)=H(X_{d+1}|X_{2}^{d},Y_{1})\leq R_{\infty}(0)=H(\mathcal{X}),\label{eq:entropy rate in}
\end{equation}
since $R_{\infty}(0)$ is the required rate without side information.
This result is proved by the chain of inequalities $H(X_{d+1}|X_{2}^{d},Y_{1})\leq H(X_{d+1}|X_{1}^{d})\leq H(\mathcal{X}),$
where the first inequality follows by the data processing inequality
and the second by conditioning reduces entropy. However, inequality
(\ref{eq:entropy rate in}) may not be strict, and thus side information
may not be useful. A first example is the case where $X_{i}$ is an
i.i.d. process, which is obtained by making $q(x|y)$ independent
of $y$. As another example, consider the setting of source coding
with feedforward \cite{weissman03,pradhan}, i.e., $X_{i}=Y_{i}$.
In this case, our assumption (\ref{eq:hidden Markov}) entails that
$X^{n}$ is a Markov chain, and we have $R_{d}(0)=H(X_{d+1}|X_{1}^{d})=H(X_{2}|X_{1})=H(\mathcal{X})$
for $d\geq1$. Therefore, delayed feedforward (with $d\geq1)$ is
not useful for the lossless compression of Markov chains, as already
shown in \cite{weissman03}. This conclusion need not hold for lossy
compression (i.e., for $D_{1}>0$) \cite{weissman03} (see also Sec.
\ref{sub:Binary-Hidden-Markov}). \end{rem}

\begin{rem}If $X^{n},Y^{n}$ are general jointly stationary and ergodic
processes (and not necessarily stationary ergodic hidden Markov models),
one can adapt in a straightforward way the proofs of Appendix \ref{sub:Proof-of-Converse}
and Sec. \ref{sub:Proof-of-Achievability}, and conclude that the
rate distortion function can be written as
\begin{equation}
R_{d}(0)=\lim_{n\rightarrow\infty}\frac{1}{n}H(X^{n}||Y^{n-d}),\label{eq:directed info}
\end{equation}
where $H(X^{n}||Y^{n-d})$ is the causally conditioned entropy $H(X^{n}||Y^{n-d})=\sum_{i=1}^{n}H(X_{i}|X^{i-1}Y^{i-d})$
(see, e.g., \cite{Permuter Kim})%
\footnote{The limit exists because the sequence is non-increasing and bounded
below. %
}. Comparing (\ref{eq:directed info}) with the rate $R_{\infty}(0)=H(\mathcal{X})$
necessary in the absence of any side information, we conclude that
the reduction in the compression rate obtained by leveraging delayed
side information at the decoder, when side information is known at
the encoder, is given for stationary and ergodic processes by
\begin{equation}
R_{\infty}(0)-R_{d}(D)=\lim_{n\rightarrow\infty}\frac{1}{n}I(Y^{n-d}\rightarrow X^{n}).\label{eq:directed mut info}
\end{equation}
In (\ref{eq:directed mut info}), we have used the definition of \emph{directed}
mutual information $I(Y^{n-d}\rightarrow X^{n})=H(X^{n})-H(X^{n}||Y^{n-d})$
(see, e.g., \cite{Permuter Kim}). Note that the rate gain (\ref{eq:directed mut info})
complements the results given in \cite{Permuter Kim} on the interpretation
of the directed mutual information (see also next remark). \end{rem}

\begin{rem}\label{rem:Consider-a-variable-length}Consider a \emph{variable-length}
(strictly) lossless source code that operates symbol by symbol such
that, for every symbol $i\in[1,n]$, it outputs a string of bits $M_{i}(X^{i},Y^{i-d}),$
which is a function of $X^{i}$ and $Y^{i-d}$. Encoding is constrained
so that the code $M_{i}(x^{i},y^{i-d})$ for each ($x^{i},y^{i-d}$)
is prefix-free. The decoder, based on delayed side information, can
then uniquely decode each codeword $M_{i}(x^{i},y^{i-d})$ as soon
as it is received. Following the considerations in \cite[Sec. IV]{Permuter Kim},
it is easy to verify that rate $R_{d}(0)$ (and, more generally, (\ref{eq:directed info}))
is also the infimum of the average rate in bits/source symbol required
by such code. Moreover, it is possible to construct universal context-based
compression strategies by adapting the approach in \cite{Willems}.
\end{rem}

We refer to Sec. \ref{sec:Examples} for some examples that further
illustrate some implications of Proposition 1.

\begin{figure}
\begin{centering}
\includegraphics[width=4.5in]{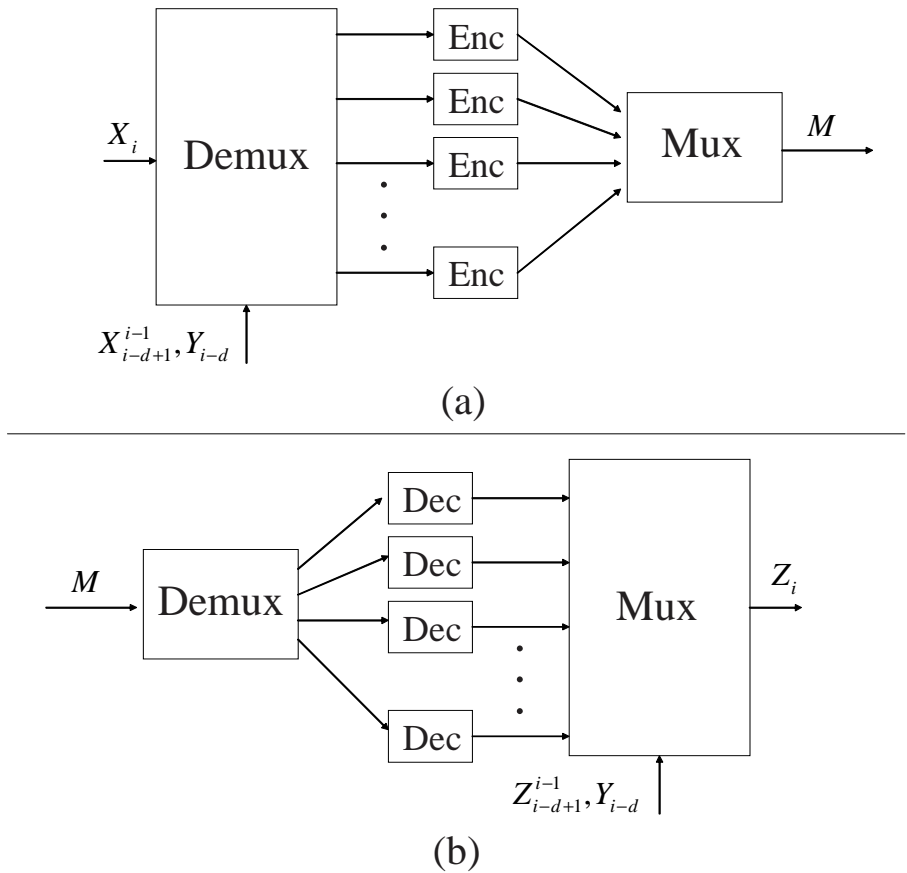}
\par\end{centering}

\caption{A block diagram for encoder (a) and decoder (b) used in the proof
of achievability of Proposition 1.}

\label{figmux}
\end{figure}

\begin{figure}
\begin{centering}
\centering\includegraphics[width=3.6in]{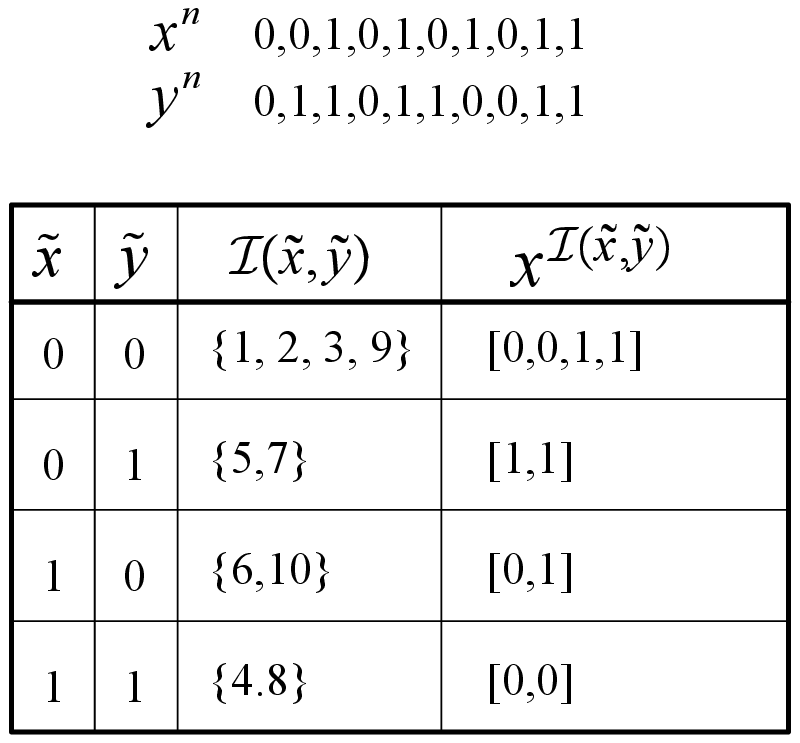}
\par\end{centering}

\caption{An example that illustrates the operations of the {}``Demux'' block
of the encoder used for the achievability proof of Proposition 1,
as shown in Fig. \ref{figmux}, for $d=2$ (symbols corresponding
to out-of-range indices are set to zero).}

\label{figillustration}
\end{figure}

\subsection{Proof of Achievability for Proposition 1\label{sub:Proof-of-Achievability}}
\begin{IEEEproof}
(Achievability) Here we propose a coding scheme that achieves rate
(\ref{lossless}). The basic idea is a non-trivial extension of the
approach discussed in \cite[Remark 3, p. 5227]{El Gamal} and is described
as follows. A block diagram is shown in Fig. \ref{figmux} for encoder
(Fig. \ref{figmux}-(a)) and decoder (Fig. \ref{figmux}-(b)).

We first describe the \emph{encoder,} which is illustrated in Fig.
\ref{figmux}-(a). To encode sequences $(x^{n},y^{n})\in(\mathcal{X}^{n}\times\mathcal{Y}^{n}),$
we first partition the interval $[1,n]$ into $|\mathcal{X}|^{d-1}|\mathcal{Y}|$
subintervals, which we denote as $\mathcal{I}(\tilde{x}^{d-1},\tilde{y})\subseteq\lbrack1,n]$,
for all $\tilde{x}^{d-1}\in\mathcal{X}^{d-1}$ and $\tilde{y}\in\mathcal{Y}$.
Every such subinterval $\mathcal{I}(\tilde{x}^{d-1},\tilde{y})$ is
defined as
\begin{equation}
\mathcal{I}(\tilde{x}^{d-1},\tilde{y})=\{i\text{: }i\in\lbrack1,n]\text{ and }y_{i-d}=\tilde{y},\text{ }x_{i-d+1}^{i-1}=\tilde{x}^{d-1}\}.\label{eq:I(xy)}
\end{equation}
In words, the subinterval $\mathcal{I}(\tilde{x}^{d-1},\tilde{y})$
contains all symbol indices $i$ such that the corresponding delayed
side information available at the decoder is $y_{i-d}=\tilde{y}$
and the previous $d-1$ samples in $x^{n}$ are $x_{i-d+1}^{i-1}=\tilde{x}^{d-1}$.
We refer to the value of the tuple ($y_{i-d},x_{i-d+1}^{i-1}$) as
the \emph{context} of sample $x_{i}$.%
\footnote{For the feedforward case $X_{i}=Y_{i}$, this definition of context
is consistent with the conventional one given in \cite{Rissanen}
when specialized to Markov processes. See also Remark \ref{rem:Consider-a-variable-length}.%
} For the out-of-range indices $i\in[-d+1,0]$, one can assume arbitrary
values for $x_{i}\in\mathcal{X}$ and $y_{i}\in\mathcal{Y}$, which
are also shared with the decoder once and for all. Note that $\bigcup_{\tilde{x}^{d-1}\in\mathcal{X}^{d-1},\mbox{ }\tilde{y}\in\mathcal{Y}}\mathcal{I}(\tilde{x}^{d-1},\tilde{y})=[1,n]$.
Fig. \ref{figillustration} illustrates the definitions at hand for
$d=2$.

As a result of the partition described above, the encoder {}``demultiplexes''
sequence $x^{n}$ into $|\mathcal{X}|^{d-1}|\mathcal{Y}|$ sequences
$x^{\mathcal{I}(\tilde{x}^{d-1},\tilde{y})}$, one for each possible
context ($\tilde{x}^{d-1},\tilde{y})\in\mathcal{X}^{d-1}\mathcal{\times Y}$.
This demultiplexing operation, which is controlled by the previous
values of source and side information, is performed in Fig. \ref{figmux}-(a)
by the block labelled as {}``Demux'', and an example of its operation
is shown in Fig. \ref{figillustration}. By the ergodicity of process
$X_{i}$ and $Y_{i}$, for every $\epsilon>0$ and all sufficiently
large $n$, the length of any sequence $x^{\mathcal{I}(\tilde{x}^{d-1},\tilde{y})}$
is guaranteed to be less than $np_{Y_{1}X_{2},...,X_{d}}(\tilde{y},\tilde{x}^{d-1})+\epsilon$
symbols with probability arbitrarily close to one. This because the
length $|\mathcal{I}(\tilde{x}^{d-1},\tilde{y})|$ of the sequence
$x^{\mathcal{I}(\tilde{x}^{d-1},\tilde{y})}$ equals the number of
occurrences of the context ($y_{i-d}=\tilde{y},\text{ }x_{i-d+1}^{i-1}=\tilde{x}^{d-1}$)
and by Birkhoff's ergodic theorem (see \cite[Sec. 16.8]{Cover}).
In particular, for any $\epsilon>0$ we can find an $n$ such that
\begin{equation}
\Pr[\mathcal{E}_{1}(\tilde{y},\tilde{x}^{d-1})]\leq\frac{\epsilon}{2|\mathcal{X}|^{d-1}|\mathcal{Y}|},\label{eq:error_1}
\end{equation}
where we have defined the {}``error'' event
\begin{equation}
\mathcal{E}_{1}(\tilde{y},\tilde{x}^{d-1})=\{|\mathcal{I}(\tilde{x}^{d-1},\tilde{y})|>np_{Y_{1}X_{2},...,X_{d}}(\tilde{y},\tilde{x}^{d-1})+\epsilon\}.
\end{equation}

Each sequence $x^{\mathcal{I}(\tilde{x}^{d-1},\tilde{y})}$ is encoded
by a separate encoder, labelled as {}``Enc'' in Fig. \ref{figmux}-(a).
In case the cardinality $|\mathcal{I}(\tilde{x}^{d-1},\tilde{y})|$
does not exceed $np_{Y_{1}X_{2},...,X_{d}}(\tilde{y},\tilde{x}^{d-1})+\epsilon$
(i.e., the {}``error'' event $\mathcal{E}_{1}(\tilde{y},\tilde{x}^{d-1})$
does not occur), the encoder compresses sequence $x^{\mathcal{I}(\tilde{x}^{d-1},\tilde{y})}$
using an entropy encoder, as explained below. If the cardinality condition
is instead not satisfied (i.e., $\mathcal{E}_{1}(\tilde{y},\tilde{x}^{d-1})$
is realized), then an arbitrary bit sequence of length $L_{\epsilon}(\tilde{y},\tilde{x}^{d-1})$,
to be specified below, is selected by the encoder {}``Enc''.

The entropy encoder can be implemented in different ways, e.g., using
typicality or Huffman coding (see, e.g., \cite{Cover}). Here we consider
a typicality-based encoder. Note that the entries $X_{i}$ of each
sequence $X^{\mathcal{I}(\tilde{x}^{d-1},\tilde{y})}$ are i.i.d.
with distribution $p_{X_{d+1}|Y_{1}X_{2},...,X_{d}}(\cdot|\tilde{y},\tilde{x}^{d-1})$,
since conditioning on the context $\{y_{i-d}=\tilde{y},\text{ }x_{i-d+1}^{i-1}=\tilde{x}^{d-1}\}$
makes the random variables $X_{i}$ independent. As it is standard
practice, the entropy encoder assigns a distinct label to all $\epsilon$-typical
sequences $\mathcal{T}_{\epsilon}(p_{X_{d+1}|Y_{1}X_{2},...,X_{d}}(\cdot|\tilde{y},\tilde{x}^{d-1}))$
with respect to such distribution, and an arbitrary label to non-typical
sequences. From the Asymptotic Equipartion Property (AEP), we can
choose $n$ sufficiently large so that (see, e.g., \cite{El Gamal Kim})
\begin{equation}
\Pr[\mathcal{E}_{2}(\tilde{y},\tilde{x}^{d-1})]\leq\frac{\epsilon}{2|\mathcal{X}|^{d-1}|\mathcal{Y}|},\label{eq:error_2}
\end{equation}
where we have defined the {}``error'' event
\begin{equation}
\mathcal{E}_{2}(\tilde{y},\tilde{x}^{d-1})=\{X^{\mathcal{I}(\tilde{x}^{d-1},\tilde{y})}\notin\mathcal{T}_{\epsilon}(p_{X_{d+1}|Y_{1}X_{2},...,X_{d}}(\cdot|\tilde{y},\tilde{x}^{d-1}))\}.
\end{equation}
Moreover, by the AEP, a rate in bits per source symbol of $H(X_{d+1}|X_{2}^{d}=\tilde{x}^{d-1},Y_{1}=\tilde{y})+\epsilon$
is sufficient for the entropy encoder to label all $\epsilon$-typical
sequences.

From the discussion above, it follows that the proposed scheme encodes
each sequence $x^{\mathcal{I}(\tilde{x}^{d-1},\tilde{y})}$ with $L_{\epsilon}(\tilde{y},\tilde{x}^{d-1})=np_{Y_{1}X_{2},...,X_{d}}(\tilde{y},\tilde{x}^{d-1})H(X_{d+1}|X_{2}^{d}=\tilde{x}^{d-1},Y_{1}=\tilde{y})+n\delta(\epsilon)$
bits. By concatenating the descriptions of all the $|\mathcal{X}|^{d-1}|\mathcal{Y}|$
sequences $x^{\mathcal{I}(\tilde{x}^{d-1},\tilde{y}_{1})}$, we thus
obtain that the overall rate $R$ of message $M$ for the scheme at
hand is $H(X_{d+1}|X_{2}^{d-1},Y_{1})+\delta(\epsilon)$. The concatenation
of the labels output by each entropy encoder is represented in Fig.
\ref{figmux}-(a) by the block {}``Mux''. We emphasize that encoder
and decoder agree a priori on the order in which the descriptions
of the different subsequences are concatenated. For instance, with
reference to the example in Fig. \ref{figillustration} (with $d=2$),
message $M$ can contain first the description of the sequence corresponding
to $(\tilde{x},\tilde{y})=(0,0)$, then $(\tilde{x},\tilde{y})=(0,1)$,
etc.

We now describe the \emph{decoder,} which is illustrated in Fig. \ref{figmux}-(b).
By undoing the multiplexing operation just described, the decoder,
from the message $M$, can recover the individual sequences $x^{\mathcal{I}(\tilde{x}^{d-1},\tilde{y})}$
through a simple demultiplexing operation for all contexts $(\tilde{x}^{d-1},\tilde{y})\in\mathcal{X}^{d-1}\times\mathcal{Y}$.
This operation is represented by block {}``Demux'' in Fig. \ref{figmux}-(b).
To be precise, this demultiplexing is possible, unless the encoding
{}``error'' event
\begin{equation}
\mathcal{E}=\bigcup_{\tilde{x}^{d-1}\in\mathcal{X}^{d-1},\mbox{ }\tilde{y}\in\mathcal{Y}}\{\mathcal{E}_{1}(\tilde{y},\tilde{x}^{d-1})\cup\mathcal{E}_{2}(\tilde{y},\tilde{x}^{d-1})\}
\end{equation}
takes place. In fact, occurrence of the {}``error'' event $\mathcal{E}$
implies that some of the sequences $x^{\mathcal{I}(\tilde{x}^{d-1},\tilde{y})}$
was not correctly encoded and hence cannot be recovered at the decoder.
The effect of such errors will be accounted for below.

Assume now that no error has taken place in the encoding. While the
individual sequences $x^{\mathcal{I}(\tilde{x}^{d-1},\tilde{y})}$
can be recovered through the discussed demultiplexing operation, this
does not imply that the decoder is also able to recover the original
sequence $x^{n}$. In fact, that decoder does not know a priori the
partition $\{\mathcal{I}(\tilde{x}^{d-1},\tilde{y})$: $\tilde{x}^{d-1}\in\mathcal{X}^{d-1}$
and $\tilde{y}\in\mathcal{Y}\}$ of the interval $[1,n]$ and thus
cannot reorder the elements of sequences $x^{\mathcal{I}(\tilde{x}^{d-1},\tilde{y})}$
to produce $x^{n}$. Recall, moreover, that such re-ordering operation
should be done in a causal fashion following the decoding rule (\ref{eq:decoder}).

We now argue that the re-ordering mentioned above is in fact possible
using a decoding rule that complies with (\ref{eq:decoder}) via a
multiplexing block controlled by the previous estimates of the source
samples (block {}``Mux'' in Fig. \ref{figmux}-(b)). In fact, note
that at time $i$, the decoder knows $Y_{i-d}$ and the previously
decoded $X^{i-1}$ and can thus identify the subinterval $\mathcal{I}(\tilde{x}^{d-1},\tilde{y})$
to which the current symbol $X_{i}$ belongs. This symbol can be then
immediately read as the next yet-to-be-read symbol from the corresponding
sequence $x^{\mathcal{I}(\tilde{x}^{d-1},\tilde{y})}$. Note that
for the first $d$ symbols, the decoder uses the values for $x_{i}$
and $y_{i}$ at the out-of-range indices $i$ that were agreed upon
with the encoder (see above). In conclusion, we remark that the scheme
described above, by choosing $\epsilon$ small enough and $n$ large
enough, is able to satisfy the constraint (\ref{dist constraints-1})
to any desired accuracy. We also note that the controlled multiplexing/demultiplexing
operation used in the proof is reminiscent of the scheme proposed
in \cite{Goldsmith} for transmission on fading channels with side
information at the transmitter and receiver.

We finally need to study the effect of errors. Given the choices made
above, we have that the probability of an encoding error is
\begin{align}
\Pr[\mathcal{E}] & \leq\sum_{\tilde{x}^{d-1}\in\mathcal{X}^{d-1},\mbox{ }\tilde{y}\in\mathcal{Y}}\Pr[\mathcal{E}_{1}(\tilde{y},\tilde{x}^{d-1})]+\Pr[\mathcal{E}_{2}(\tilde{y},\tilde{x}^{d-1})]\leq\epsilon,
\end{align}
where the first inequality follows from the union bound and the second
from (\ref{eq:error_1}) and (\ref{eq:error_2}). This implies that
the distortion in (\ref{dist constraints-1}) is upper bounded by
$\epsilon$ as desired. In fact, from the definition of encoder and
decoder given above, we can conclude that $\Pr[X^{n}\neq Z_{1}^{n}]=\Pr[\mathcal{E}]\leq\epsilon$,
where we recall that $Z_{1}^{n}$ is the sequence reconstructed at
the decoder. Moreover, the following inequality holds in general
\begin{equation}
\Pr[X^{n}\neq Z_{1}^{n}]\geq\frac{1}{n}\sum\limits _{i=1}^{n}\Pr[X_{i}\neq Z_{1i}].\label{eq:ineqperr}
\end{equation}
Therefore, we have $\frac{1}{n}{\textstyle {\textstyle {\displaystyle \sum\limits _{i=1}^{n}}}}\Pr[X_{i}\neq Z_{1i}]\leq\epsilon$,
which concludes the proof.
\end{IEEEproof}
\begin{rem} An alternative proof of achievability can be given by
using the idea of codetrees and extending the notions of typicality
introduced in \cite{pradhan}. The proof discussed above is based
on a conceptually and algorithmically simpler approach, albeit its
applicability is limited to lossless compression (see next subsection).\end{rem}

\begin{rem} From the inequality (\ref{eq:ineqperr}), it follows
that the optimality of the scheme above can be proved also under the
more stringent block error probability constraint (see also \cite[Sec. 3.6.4]{El Gamal Kim}).\end{rem}

\subsection{Lossy Compression}

Here, we obtain a characterization of the rate-distortion function
$R_{d}(D_{1})$, for $d=0$ and $d=1$. The proof follows as a special
case of that of Proposition \ref{pro:For-any-delay} to be discussed
in the next section, and is based on similar arguments as for Proposition
\ref{prop:For-any-delay 1}.
\begin{prop}
\label{cor:For-any-delay}For any delay $d\geq0$ and distortion $D_{1}$,
the following rate is achievable for the setting of Fig. \ref{fig0}
\begin{equation}
R_{d}^{(a)}(D_{1})=\min I(XY;Z_{1}|Y_{d}),\label{eq:Rda}
\end{equation}
 with mutual informations evaluated with respect to the joint distribution
\begin{equation}
p(x,y,y_{d},z_{1})=\pi(y_{d})w_{d}(y|y_{d})q(x|y)p(z_{1}|x,y,y_{d}),\label{eq:distr R1d1d2-1-1}
\end{equation}
 and where minimization is done over all conditional distributions
$p(z_{1}|x,y,y_{d})$ such that
\begin{equation}
\mathrm{E}[d_{1}(X,Y,Z_{1})]\leq D_{1}.\label{eq:dist const R1d1d2-1-1}
\end{equation}
 Moreover, rate (\ref{eq:Rda})-(\ref{eq:dist const R1d1d2-1-1})
is the rate-distortion function, i.e., \textup{$R_{d}^{(a)}(D_{1})=R_{d}(D_{1})$},
for $d=0$ and $d=1$.
\end{prop}
\begin{rem} The optimality of the conditional codebook strategy for
lossless compression shown in Proposition \ref{prop:For-any-delay 1}
hinges on the following fact: conditioned on the context (\emph{$Y_{i-d},X_{i-d+1},$}
$\text{\ldots},X_{i-1}$), the samples \emph{$X_{i}$} are independent
of the past samples $X^{i-1}$ by the hidden Markov model assumption.
Recall that the fact that the decoder has available the past source
samples ($X_{i-d+1},$$\text{\ldots},X_{i-1}$) since its estimates
are correct with high probability. Due to this independence property,
and to the availability of the side information also at the encoder,
the latter need not use \textquotedblleft{}multi-letter\textquotedblright{}
compression codes and can instead use simple \textquotedblleft{}single-letter\textquotedblright{}
entropy codes conditioned on the values of (\emph{$Y_{i-d},X_{i-d+1},$}$\text{\ldots},X_{i-1}$)
without loss of optimality. In the lossy case considered in Proposition
\ref{cor:For-any-delay}, instead, even for the point-to-point model,
the independence condition discussed above does not hold for delays
$d$ strictly larger than 1. In fact, at each time $i$, the decoder
has available the delayed side information $Y^{i-d}$ only, conditioned
on which the source samples $X_{i}$ are not independent of the past
samples $X^{i-1}$. But, for $d=1$, the independence condition at
hand does apply and thus the optimality of \textquotedblleft{}single-letter\textquotedblright{}
codes can be proved as done in Proposition \ref{cor:For-any-delay}.
\end{rem}

\section{When the Side Information May Be Delayed\label{sec:Lossy-Source-Coding}}

In this section, we consider the problem of lossy compression for
the set-up of Fig. \ref{fig2}. Note that the asymptotically lossless
case follows from Proposition \ref{prop:For-any-delay 1}, since,
in order to guarantee lossless reconstruction also at the decoder
with delayed side information, rate $R$ must satisfy the conditions
in Proposition \ref{prop:For-any-delay 1}. Here, we obtain an achievable
rate region $\mathcal{R}_{d}^{(a)}(D_{1},D_{2})\mathcal{\subseteq R}_{d}(D_{1},D_{2})$
for all delays $d\geq0$ for the model in Fig. \ref{fig2}, and show
that such region coincides with the rate-distortion region, i.e.,
$\mathcal{R}_{d}^{(a)}(D_{1},D_{2})\mathcal{=R}_{d}(D_{1},D_{2})$,
for $d=0$ and $d=1$.

To streamline the discussion, we start by consider the special case
where $\Delta R=0$ and obtain a characterization of the rate-distortion
function $R_{d}(D_{1},D_{2})$ for $d=0$ and $d=1$.
\begin{prop}
\label{pro:2}For any delay $d\geq0$ and distortion pair $(D_{1},D_{2})$,
the following rate is achievable for the set-up of Fig. \ref{fig2}
with $\Delta R=0$
\begin{align}
R_{d}^{(a)}(D_{1},D_{2}) & =\min I(XY;Z_{1}|Y_{d})+I(X;Z_{2}|YY_{d}Z_{1})\label{eq:R1d1d2a}\\
 & =\min I(Y;Z_{1}|Y_{d})+I(X;Z_{1}Z_{2}|YY_{d}),\label{eq:R1d1d2-1}
\end{align}
 with mutual informations evaluated with respect to the joint distribution
\begin{equation}
p(x,y,y_{d},z_{1},z_{2})=\pi(y_{d})w_{d}(y|y_{d})q(x|y)p(z_{1},z_{2}|x,y,y_{d}),\label{eq:distr R1d1d2-1}
\end{equation}
 and where minimization is done over all conditional distributions
$p(z_{1},z_{2}|x,y,y_{d})$ such that
\begin{equation}
\mathrm{E}[d_{j}(X,Y,Z_{j})]\leq D_{j}\text{, for }j=1,2.\label{eq:dist const R1d1d2-1}
\end{equation}
 Moreover, rate (\ref{eq:R1d1d2a})-(\ref{eq:R1d1d2-1}) is the rate-distortion
function, i.e., \textup{$R_{d}^{(a)}(D_{1},D_{2})=R_{d}(D_{1},D_{2})$},
for $d=0$ and $d=1$.
\end{prop}
\begin{rem} \label{rem:rate inter-1}Rate (\ref{eq:R1d1d2a}) can
be easily interpreted in terms of achievability. To this end, we remark
that variable $Y_{d}$ plays the role of the delayed side information
$Y^{i-d}$ at decoder 1. The coding scheme achieving rate (\ref{eq:R1d1d2a})
operates in two successive phases. In the first phase, the encoder
encodes the reconstruction sequence $Z_{1}^{n}$ for decoder 1. Since
decoder 1 has available delayed side information, using a strategy
similar to the one discussed in Sec. \ref{sub:Proof-of-Achievability},
this operation requires $I(XY;Z_{1}|Y_{d})$ bits per source sample,
as further detailed in Sec. \ref{sub:Proof-of-Achievability-2}. Note
that decoder 2 is able to recover $Z_{1}^{n}$ as well, since decoder
2 has available side information $Y^{i}$, and thus also the delayed
side information $Y^{i-d}$. In the second phase, the reconstruction
sequence $Z_{2}^{n}$ for decoder 2 is encoded. Given the side information
available at decoder 2, this operation requires rate $I(X;Z_{2}|YY_{d}Z_{1})$,
using again an approach similar to the one discussed in Sec. \ref{sub:Proof-of-Achievability}.
The converse proof is in Appendix \ref{sub:Proof-of-Converse-2}.\end{rem}

\begin{rem}\label{rem:For-memoryless-sources-1}For memoryless sources
$X^{n}$ and $Y^{n}$, obtained by setting the transition probability
$w_{1}(y_{i}|y^{i-1})$ to be independent of $y^{i-1}$, it can be
seen that the achievable rate (\ref{eq:R1d1d2a})-(\ref{eq:R1d1d2-1})
is the rate-distortion function for the scenario of Fig. \ref{fig2}
with $\Delta R=0$ \emph{for all delays} $d\geq0$. This observation
extends Lemma 1 to the more general set-up of Fig. \ref{fig2} with
$\Delta R=0$. To see this, note that for $d\geq1$, rate (\ref{eq:R1d1d2a})-(\ref{eq:R1d1d2-1})
is given by
\begin{equation}
R_{d}^{(a)}(D_{1},D_{2})=\min I(XY;Z_{1})+I(X;Z_{2}|YZ_{1}),\label{eq:rem memoryless-1}
\end{equation}
with mutual informations evaluated with respect to the joint distribution
\begin{equation}
p(x,y,y_{d},z_{1},z_{2})=\pi(y)q(x|y)p(z_{1},z_{2}|x,y),\label{eq:distr R1d1d2-12-1}
\end{equation}
and where minimization is done over all conditional distributions
$p(z_{1},z_{2}|x,y,y_{d})$ such that the distortion constraints (\ref{eq:dist const R1d1d2-1})
are satisfied. Rate (\ref{eq:rem memoryless-1}) recovers the rate-distortion
function derived by \cite{Kaspi} for the case where decoder 1 has
\emph{no} side information. Therefore, rate (\ref{eq:rem memoryless-1})
is achievable even without any state information at decoder 1. We
then conclude that\emph{ }delayed side information is not useful for
memoryless sources\emph{. }Note also that \cite{Kaspi} assumes non-causal
availability of the side information at decoder 2. The equality of
the rate derived in \cite{Kaspi} and the one in Proposition \ref{pro:2}
thus demonstrates that causal and non-causal side information lead
to the same performance in terms of rate-distortion function. \end{rem}

\begin{rem}While (\ref{eq:R1d1d2a}) is easier to interpret in terms
of achievability as done in Remark \ref{rem:rate inter-1}, the equivalent
expression (\ref{eq:R1d1d2-1}) highlights the rate loss due to the
possible delay of the side information. In fact, the mutual information
$I(X;Z_{1}Z_{2}|YY_{d})$ accounts for the rate that would be needed
to convey both $Z_{1}^{n}$ and $Z_{2}^{n}$ only to decoder 2, which
has non-delayed side information. Therefore, the additional term $I(Y;Z_{1}|Y_{d})$
can be interpreted as the extra rate that needs to be expended to
enable transmission of $Z_{1}^{n}$ also to decoder 1, which has delayed
side information. \end{rem}

We now consider the general model in Fig. \ref{fig2}.
\begin{prop}
\label{pro:For-any-delay}For any delay $d\geq0$ and any distortion
pair ($D_{1},D_{2}$), define $\mathcal{R}_{d}^{(a)}(D_{1},D_{2})$
as the union of all rate pairs ($R,\Delta R$) that satisfy
\begin{align}
R & \geq I(Y;Z_{1}|Y_{d})+I(X;Z_{1}U|YY_{d})\label{eq:ineq fig1}\\
R+\Delta R & \geq I(Y;Z_{1}|Y_{d})+I(X;Z_{1}Z_{2}U|YY_{d})\label{eq:ineq fig2}
\end{align}
 for some joint distribution
\begin{equation}
p(x,y,y_{d},u,z_{1},z_{2})=\pi(y_{d})w_{d}(y|y_{d})q(x|y)p(z_{1},z_{2},u|x,y,y_{d})\label{sr:pmf}
\end{equation}
 where minimization is done over all conditional distributions $p(z_{1},z_{2},u|x,y,y_{d})$
such that
\begin{equation}
\mathrm{E}[d_{j}(X,Y,Z_{j})]\leq D_{j}\text{, for }j=1,2.
\end{equation}
 We have that
\begin{equation}
\mathcal{R}_{d}^{(a)}(D_{1},D_{2})\mathcal{\subseteq R}_{d}(D_{1},D_{2})\label{eq:ach region fig2}
\end{equation}
 for any $d\geq0$. Moreover, equation (\ref{eq:ach region fig2})
holds with equality, and thus $\mathcal{R}_{d}^{(a)}(D_{1},D_{2})$
is the rate-distortion region, for $d=0$ and $d=1$.
\end{prop}
\begin{rem} Let us interpret the rate region $\mathcal{R}_{d}^{(a)}(D_{1},D_{2})$
in terms of achievability. First, from Remark \ref{rem:rate inter-1},
we observe that (\ref{eq:ineq fig1}) is the rate necessary to convey
$Z_{1}^{n}$ to both decoder 1 and decoder 2, and an auxiliary codeword
$U^{n}$ only to decoder 2. This auxiliary codeword $U^{n}$ carries
information to decoder 2 that is then refined via message $M_{\Delta}.$
In particular, rewriting (\ref{eq:ineq fig2}) as $R+\Delta R\geq I(Y;Z_{1}|Y_{d})+I(X;Z_{1}U|YY_{d})+I(X;Z_{2}|YY_{d}UZ_{1})$,
by comparison with (\ref{eq:ineq fig1}), we see that the extra rate
$I(X;Z_{2}|YY_{d}UZ_{1})$ is needed to transmit sequence $Z_{2}^{n}$
to decoder 2, thus refining the information available therein due
to message $M$.%
\footnote{Note that such rate can be encoded in both messages $M$ and $M_{\Delta}$,
which leads to the sum-rate constraint (\ref{eq:ineq fig2}).%
}\end{rem}

\begin{rem} The considerations in Remark \ref{rem:For-memoryless-sources-1}
can be also easily extended to the scenario of Proposition \ref{pro:For-any-delay}
with $\Delta R\geq0$. \end{rem}

\subsection{Proof of Achievability of Proposition \ref{pro:2} and Proposition
\ref{pro:For-any-delay} \label{sub:Proof-of-Achievability-2}}
\begin{figure}
\begin{centering}
\includegraphics[width=3.6in]{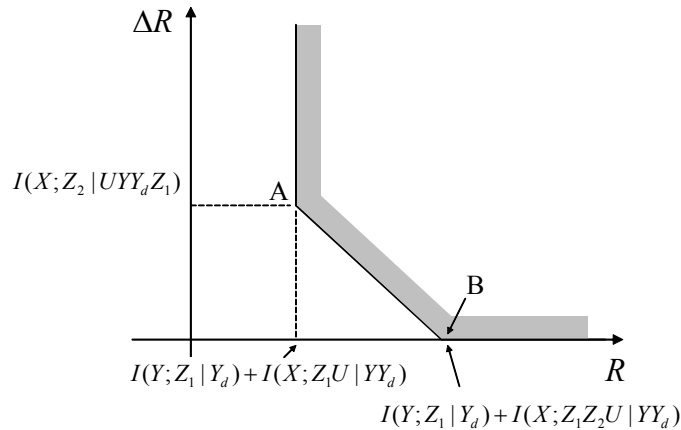}
\par\end{centering}

\caption{Achievable rate region used in the proof of Proposition \ref{pro:For-any-delay}.}

\label{figregion}
\end{figure}

\begin{IEEEproof}
(Achievability) We first prove achievability of rate (\ref{eq:R1d1d2a})
in Proposition \ref{pro:2}. The proof extends the ideas discussed
in Sec. \ref{sub:Proof-of-Achievability}, to which we refer for details.
In particular, here we do not detail the calculations of the encoding
{}``error'' events and distortion levels, as they follow in the
same way as in Sec. \ref{sub:Proof-of-Achievability}. To encode sequence
($x^{n},y^{n}$), the encoder partitions the interval $[1,n]$ into
$|\mathcal{Y}|$ subintervals, namely $\mathcal{I}(\tilde{y})$ for
each $\tilde{y}\in\mathcal{Y}$, so that (cf. (\ref{eq:I(xy)}))
\begin{equation}
\mathcal{I}(\tilde{y})=\{i\text{: }i\in\lbrack1,n]\text{ and }y_{i-d}=\tilde{y}\}.\label{eq:I(xy)-1}
\end{equation}
 Similar to Sec. \ref{sub:Proof-of-Achievability}, a different compression
codebook is used for each such interval $\mathcal{I}(\tilde{y})$,
and thus for each pair of {}``demultiplexed'' subsequences $(x^{\mathcal{I}(\tilde{y})},y^{\mathcal{I}(\tilde{y})})$.
The compression of each pair of sequences $(x^{\mathcal{I}(\tilde{y})},y^{\mathcal{I}(\tilde{y})})$
is based on a test channel $p(z_{1}|x,y,\tilde{y}).$ Specifically,
the corresponding codewords $Z_{1}^{n}$ are generated i.i.d. according
to the marginal distribution $\sum_{(x,y)\in\mathcal{Y}}p(z_{1}|x,y,\tilde{y})$
$w_{1}(y|\tilde{y})q(x|y)$ and compression is done based on standard
joint typicality arguments. By the covering lemma \cite{El Gamal Kim},
compression of sequences $(X^{\mathcal{I}(\tilde{y})},Y^{\mathcal{I}(\tilde{y})})$
into the corresponding reconstruction sequence $Z_{1}^{\mathcal{I}(\tilde{y})}$
requires rate $I(XY;Z_{1}|\tilde{Y}=\tilde{y})+\epsilon$ bits per
source symbol in each interval $\mathcal{I}(\tilde{y})$, and thus
an overall rate $I(XY;Z_{1}|\tilde{Y})+\epsilon$ following the same
considerations as in Sec. \ref{sub:Proof-of-Achievability}. In particular,
the encoder multiplexes the compression indices corresponding to the
$|\mathcal{Y}|$ intervals $\mathcal{I}(\tilde{y})$ to produce message
$M$. Therefore, the latter only carries information about the individual
sequences $Z_{1}^{\mathcal{I}(\tilde{y})},$ but not about the ordering
of each entry within the overall sequence $Z_{1}^{n}$.

Based on the sequence $z_{1}^{n}$ produced in the first encoding
phase described above, the encoder then performs also a finer partition
of the interval $[1,n]$ into $|\mathcal{Y}|^{2}|\mathcal{Z}_{1}|$
intervals $\mathcal{I}(\tilde{y},y,z),$ with $\tilde{y}\in\mathcal{Y},\mbox{ }y\in\mathcal{Y},\mbox{ and }z\in\mathcal{Z}$,
so that
\begin{equation}
\mathcal{I}(\tilde{y},y,z)=\{i\text{: }i\in\lbrack1,n]\text{ and }y_{i-d}=\tilde{y},\mbox{ }y_{i}=y,\mbox{\mbox{ }and }z_{i}=z\}.\label{eq:I(xy)-1-1}
\end{equation}
Compression of sequence $x^{I(\tilde{y},y,z)}$ into the corresponding
reconstruction $Z_{2}^{\mathcal{I}(\tilde{y},y,z)}$ is carried out
according to test channel $p(z_{2}|x,y,\tilde{y},z)$ as per the discussion
above, requiring an overall rate of $I(X;Z_{2}|Y\tilde{Y}Z_{1})+\epsilon.$
The compression indices for all sets $\mathcal{I}(\tilde{y},y,z)$
are concatenated in message $M$ following the compression indices
obtained from the sets $\mathcal{I}(\tilde{y})$.

Upon reception of message $M$, decoder 1 and 2 can both recover the
sequences $Z_{1}^{\mathcal{I}(\tilde{y})}$ and $Z_{2}^{\mathcal{I}(\tilde{y},y,z)}$
for all $\tilde{y}\in\mathcal{Y},$ $y\in\mathcal{Y}$ and $z\in\mathcal{Z}$
via simple demultiplexing. Moreover, following the same reasoning
as in Sec. \ref{sub:Proof-of-Achievability}, decoder 1 can reconstruct
sequence $Z_{1}^{n}$ in the correct order in a causal fashion, using
a decoder (\ref{eq:decoder}), which depends on message and delayed
side information, since the value of $Z_{1i}$ can be obtained from
sequences $Z_{1}^{\mathcal{I}(\tilde{y})}$ by knowing the value of
$Y_{i-d}.$ Similarly, decoder 2 can reorder sequence $Z_{2}^{n}$
in a causal fashion using a decoder of the form (\ref{eq:decoder2}).
This concludes the proof of achievability for Proposition \ref{pro:2}.
\hfill \QED

We now turning to the proof of achievability Proposition \ref{pro:For-any-delay}.
For a fixed distribution (\ref{sr:pmf}), we need to prove that the
rate region in Fig. \ref{figregion} is achievable. To do this, it
is enough, by standard time-sharing arguments, to prove that corner
points A and B are achievable. Corner point B corresponds to rate
pair $R=I(Y;Z_{1}|Y_{d})+I(X;Z_{1}Z_{2}U|YY_{d})$ and $\Delta R=0$.
But achievability of this region follows immediately from Proposition
\ref{pro:2} by setting $U=(UZ_{2})$ in (\ref{eq:R1d1d2a}). Instead,
corner point A corresponds to the rate pair
\begin{align}
R & =I(Y;Z_{1}|Y_{d})+I(X;Z_{1}U|YY_{d})\label{R sr}\\
\text{and }\Delta R & =I(X;Z_{2}|UYY_{d}Z_{1}).\label{DR sr}
\end{align}
This rate pair can be achieved by using a strategy similar to the
one discussed above. In this strategy, when encoding the message $M_{\Delta}$,
which is received only at decoder 2, the encoder leverages the fact
that the latter knows $Y_{i},Y_{i-d},U_{i}$ and $Z_{1i}$, by appropriately
partitioning the interval $[1,n]$ and using different test channels
in each subinterval.
\end{IEEEproof}

\section{Examples\label{sec:Examples}}

In this section, we consider two specific examples relative to the
scenario in Fig. \ref{fig0}. The first example consists of binary-alphabet
sources, while the second applies the results derived above to (continuous-alphabet)
Gaussian sources. We focus on a distortion metric of the form $\mathrm{d}_{1}(x,y,z_{1})=\mathrm{d}_{1}(x,z_{1})$
that does not depend on $y.$ In other words, the decoder is interested
in reconstructing $X^{n}$ within some distortion $D_{1}$. We note
that, under this assumption, the rate (\ref{cor:For-any-delay}) equals
the simpler expression
\begin{equation}
R_{d}^{(a)}(D_{1})=\min I(X;Z_{1}|Y_{d}),\label{eq:simpler}
\end{equation}
 with mutual informations evaluated with respect to the joint distribution
\begin{equation}
p(x,y_{d},z_{1})=\pi(y_{d})\left(\mbox{\ensuremath{\sum_{y\in\mathcal{Y}}}}w_{d}(y|y_{d})q(x|y)\right)p(z_{1}|x,y_{d}),
\end{equation}
where minimization is done over all distributions $p(z_{1}|x,y_{d})$
such that $\mathrm{E}[\mathrm{d}_{1}(X,Z_{1})]\leq D_{1}.$ Note that
this simplification is without loss of optimality because the distortion
constraint does not depend on the correlation between $Z_{1}$ and
$Y.$ Therefore, we can impose the Markov condition $Z_{1}-XY_{d}-Y$
as in (\ref{eq:simpler}) without changing the distortion, while reducing
the mutual information in (\ref{eq:Rda}).

\subsection{Binary Hidden Markov Model \label{sub:Binary-Hidden-Markov}}

In the first example, we assume that $Y_{i}$ is a binary Markov chain
with symmetric transition probabilities $w_{1}(1|0)=w_{1}(0|1)\triangleq\varepsilon$.
Therefore, we have $\pi(1)=1/2$ and $k$-step transition probabilities
$w_{k}(1|0)=w_{k}(0|1)\triangleq\varepsilon^{(k)}$, which can be
obtained recursively as $\varepsilon^{(1)}=\varepsilon$ and $\varepsilon^{(k)}=2\varepsilon^{(k-1)}(1-\varepsilon^{(k-1)})$
for $k\geq2$.%
\footnote{This follows from the standard relationship $\left[\begin{array}{cc}
1-\varepsilon^{(k)} & \varepsilon^{(k)}\\
\varepsilon^{(k)} & 1-\varepsilon^{(k)}
\end{array}\right]=\left[\begin{array}{cc}
1-\varepsilon & \varepsilon\\
\varepsilon & 1-\varepsilon
\end{array}\right]^{k}$, well known from Markov chain theory (see, e.g., \cite{Gallager}).%
} Note that this is a logistic map such that $\varepsilon^{(k)}\rightarrow1/2$
for large $k$. We also set $\varepsilon^{(0)}=0$, consistently with
the convention adopted in the rest of the paper. Finally, we assume
that
\begin{equation}
X_{i}=Y_{i}\oplus N_{i},
\end{equation}
 with {}``$\oplus$'' being the modulo-2 sum and $N_{i}$ being
i.i.d. binary variables, independent of $Y^{n}$, with $p_{N_{i}}(1)\triangleq q$,
$q\leq1/2$. We adopt the Hamming distortion $d_{1}(x,z_{1})=x\oplus z_{1}$.

\begin{figure}
\begin{centering}
\includegraphics[width=4.5in]{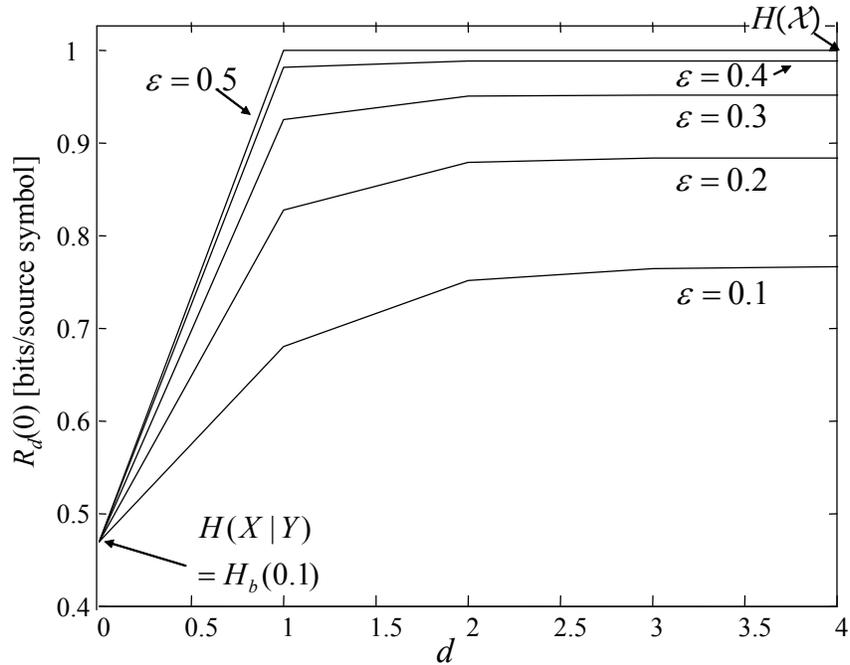}
\par\end{centering}

\caption{Minimum required rate $R_{d}(0)$ for lossless reconstruction for
the set-up of Fig. \ref{fig0} with binary sources versus delay $d$
($q=0.1$).}

\label{fig4}
\end{figure}

We start by showing in Fig. \ref{fig4} the rate $R_{d}(0)$ obtained
from Proposition 1 corresponding to zero distortion ($D_{1}=0$) versus
the delay $d$ for different values of $\varepsilon$ and for $q=0.1$.
Note that the value of $\varepsilon$ measure the {}``memory'' of
the process $Y_{i}$: For $\varepsilon$ small, the process tends
to keep its current value, while for $\varepsilon=1/2$, the values
of $Y_{i}$ are i.i.d.. For $d=0$, we have $R_{0}(0)=H(X_{1}|Y_{1})=H_{b}(q)=0.589$,
irrespective of the value of $\varepsilon$, where we have defined
the binary entropy function $H_{b}(a)=-a\log_{2}a-(1-a)\log_{2}(1-a)$.
Instead, for $d$ increasingly large, the rate $R_{d}(0)$ tends to
the entropy rate $R_{\infty}(0)=H(\mathcal{X})$. This can be calculated
numerically to arbitrary precision following \cite[Sec. 4.5]{Cover}.
Note that a larger memory, i.e., a smaller $\varepsilon$ leads to
smaller required rate $R_{d}(0)$ for all values of $d$.

\begin{figure}
\begin{centering}
\includegraphics[width=4.5in]{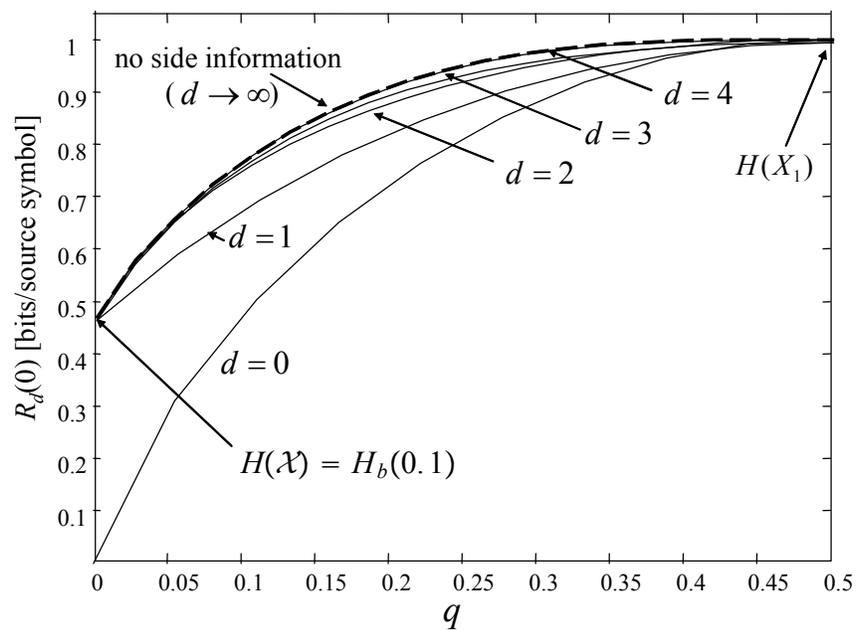}
\par\end{centering}

\caption{Minimum required rate $R_{d}(0)$ for lossless reconstruction for
the set-up of Fig. \ref{fig0} with binary sources versus parameter
$q$ ($\varepsilon=0.1$).}

\label{fig5}
\end{figure}

Fig. \ref{fig5} shows the rate $R_{d}(0)$ for $\varepsilon=0.1$
versus $q$ for different values of $d$. For reference, we also show
the performance with no side information, i.e., $R_{\infty}(0)=H(\mathcal{X})$.
For $q=1/2$, the source $X^{n}$ is i.i.d. and delayed side information
is useless in the sense that $R_{d}(0)=R_{\infty}(0)=H(X_{1})=1$
(Remark \ref{rem:Is-delayed-side}). Moreover, for $q=0$, we have
$X_{i}=Y_{i}$, so that $X_{i}$ is a Markov chain and the problem
becomes one of lossless source coding with feedforward. From Remark
\ref{rem:Is-delayed-side}, we know that delayed side information
is useless also in this case, as $R_{d}(0)=R_{\infty}(0)=H(\mathcal{X})=H_{b}(\varepsilon)=0.469$.
For intermediate values of $q$, side information is generally useful,
unless the delay $d$ is too large.

We now turn to the case where the distortion $D_{1}$ is generally
non-zero. To this end, we evaluate the achievable rate (\ref{eq:simpler})
in Appendix \ref{sec:Proof-of-()-()} obtaining
\begin{equation}
R_{d}^{(a)}(D_{1})=H_{b}(\varepsilon^{(d)}*q)-H_{b}(D_{1})\label{eq:ex1}
\end{equation}
 for
\begin{equation}
0\leq D_{1}\leq\min\{\varepsilon^{(d)}*q,1-\varepsilon^{(d)}*q\},\label{eq:ex2}
\end{equation}
 and $R_{d}^{(a)}(D_{1})=0$ otherwise. In (\ref{eq:ex1})-(\ref{eq:ex2})
we have defined $p*q\triangleq p(1-q)+(1-p)q$. Recall that rate $R_{d}^{(a)}(D_{1})$
has been proved to coincide with the rate-distortion function $R_{d}(D_{1})$
only for $d=0\mbox{ and }d=1$ (Corollary \ref{cor:For-any-delay}).

As a final remark, we use the result derived above to discuss the
advantages of delayed side information. To this end, set $q=0$ so
that $X_{i}=Y_{i}$ and the problem becomes one of source coding with
feedforward. For $d=1$, result (\ref{eq:ex1})-(\ref{eq:ex2}) recovers
the calculation in \cite[Example 2]{weissman03} (see also \cite{Naiss}),
which states that the rate-distortion function for the Markov source
$X^{n}$ at hand with feedforward ($d=1$) is
\begin{equation}
R_{1}(D)=H_{b}(\mathcal{\varepsilon})-H_{b}(D_{1})\label{slb}
\end{equation}
 for $D_{1}\leq\min(\varepsilon,1-\varepsilon)$ and $R_{1}(D_{1})=0$
otherwise. From \cite{Gray} (see also \cite{vasudevan}), it is known
that the rate-distortion function of a Markov source $X^{n}$ without
feedforward, i.e., $R_{\infty}(D_{1})$, is equal to (\ref{slb})
only for $D_{1}$ smaller than a critical value, but is otherwise
larger. This demonstrates that feedforward, unlike in the lossless
setting discussed above, can be useful in the lossy case for distortion
levels $D_{1}$ sufficiently large, as first discussed in \cite{weissman03}.

\subsection{Hidden Gauss-Markov Model\label{sub:Gaussian-Hidden-Markov}}

We now assume that $Y^{n}$ is a Gauss-Markov process with zero-mean,
power $E[Y_{i}^{2}]=1$ and correlation $\mathrm{E}[Y_{i}Y_{i+1}]=\rho$
(so that $\mathrm{E}[Y_{i}Y_{i+d}]=\rho^{d}$). Moreover, $X_{i}$
is related to $Y_{i}$ as
\begin{equation}
X_{i}=Y_{i}+N_{i},
\end{equation}
where samples $N_{i}$ are i.i.d. zero-mean Gaussian with variance
$\sigma_{N}^{2}$ and independent of $Y^{n}$. We concentrate on the
mean square error\ distortion metric $\mathrm{d}_{1}(x,z_{1})=(x-z_{1})^{2}$.
Using standard arguments, we can apply the achievable rate (\ref{eq:simpler})
to the setting at hand, although the result was derived for discrete
alphabet (see \cite[Ch. 3.8]{El Gamal Kim}). By doing so, as shown
in Appendix \ref{sec:Proof-of-()}, we get that the following rate
is achievable for $d\geq0$
\begin{equation}
R_{d}^{(a)}(D_{1})=\frac{1}{2}\log_{2}\left(\frac{1-\rho^{2d}+\sigma_{N}^{2}}{D_{1}}\right)\label{eq:gauss r}
\end{equation}
 if $0\leq D_{1}\leq1-\rho^{2d}+\sigma_{N}^{2}$ and $R_{d}^{(a)}(D_{1})=0$
otherwise. As also discussed above, this rate coincides with the rate-distortion
function for $d=0$ and $d=1$.

Similar to the discussion in the previous section for a binary hidden
Markov model, we remark that for $\sigma_{N}^{2}=0$, the problem
becomes one of lossy source coding with feedforward of a Gauss-Markov
process $X^{n}$. In this case, it is known that the rate-distortion
function without feedforward, $R_{\infty}(D_{1})$, equals $\frac{1}{2}\log_{2}\left(\frac{1-\rho^{2}}{D_{1}}\right)$
only for distortions $D_{1}$ smaller than a critical value \cite{Gray}
and is otherwise larger. By comparison with (\ref{eq:gauss r}), it
then follows that feedforward, for sufficiently large distortion levels,
can be useful in decreasing the rate-distortion function.

\section{Concluding Remarks}

The problem of compressing information sources in the presence of
delayed side information finds application in a number of scenarios
including sensor networks and prediction/denoising. A general information-theoretic
characterization of the trade-off between rate and distortion for
this problem can be generally given in terms of multi-letter expressions,
as done in \cite{pradhan directed}. Such expressions are proved by
resorting to complex achievability schemes that operate in increasingly
large blocks, and generally require involved numerical evaluations.
In this work, we have instead focused on a specific class of sources,
which evolve according to hidden Markov models, and derived single-letter
characterizations of the rate-distortion trade-off. Such characterizations
are established based on simple achievable scheme that are based on
standard {}``off-the-shelf'' compression techniques. Moreover, the
analysis has focused not only for the conventional point-to-point
setting of \cite{pradhan directed}, but also on a more general set-up
in which side information may or may not be delayed. The value of
the derived characterization is demonstrated by elaborating on two
examples, namely binary sources with Hamming distortion and Gaussian
sources with minimum mean square error distortion.

Various extensions of the results presented here are possible. For
instance, the optimal strategy for a cascade model with three nodes
in which the intermediate node has causal side information $Y^{i}$
and the end decoder has delayed side information $Y^{i-1}$ can be
identified by applying the result in Proposition \ref{pro:2} in a
manner similar to \cite{Vasudevan 2}.

\section{Acknowledgments}

The authors wish to thank Associate Editor and Reviewers for their
thoughtful comments that have helped us improve the quality of the
paper.

\appendices

\section{Proof of Converse for Proposition 1 \label{sub:Proof-of-Converse}}

For $\epsilon>0$, fix a code $(d,n,R,0,\epsilon,d_{\max})$ as defined
in Sec. \ref{sec:System-Model}. Using the definition of encoder (\ref{eq:encoder}),
we have the equalities
\begin{align}
nR & \geq H(M)=H(M)-H(M|X^{n}Y^{n})\nonumber \\
 & =I(M;X^{n}Y^{n})=H(X^{n}Y^{n})-H(X^{n}Y^{n}|M)\label{eq:app11}
\end{align}
 The first term in (\ref{eq:app11}) cam be written, using the chain
rule for entropy, as
\begin{align}
H(X^{n}Y^{n}) & =\sum_{i=1}^{d}H(X_{i}|X^{i-1})\nonumber \\
 & +\sum_{i=d+1}^{n}\left[H(Y_{i-d}|Y^{i-d-1}X^{i-1})+H(X_{i}|Y^{i-d}X^{i-1})\right]\nonumber \\
 & +\sum_{i=n-d+1}^{n}H(Y_{i}|Y^{i-1}X^{n})\nonumber \\
 & =A+\sum_{i=d+1}^{n}\left[H(Y_{i-d}|Y^{i-d-1}X^{i-1})+H(X_{i}|Y_{i-d}X_{i-d+1}^{i-1})\right]\label{eq:app12}
\end{align}
 where $A\triangleq\sum_{i=1}^{d}H(X_{i}|X^{i-1})+\sum_{i=n-d+1}^{n}H(Y_{i}|Y^{i-1}X^{n})$
is a finite constant that does not increase with $n.$ Moreover, in
the last line we have used the Markov chain $X_{i}-(Y_{i-d}X_{i-d+1}^{i-1})-Y_{1}^{i-d-1}X_{1}^{i-d}$,
which follows from (\ref{eq:hidden Markov}). The second term in (\ref{eq:app11})
can be similarly written as
\begin{align}
H(X^{n}Y^{n}|M) & =B+\sum_{i=d+1}^{n}\left[H(Y_{i-d}|Y^{i-d-1}X^{i-1}M)+H(X_{i}|Y^{i-d}X^{i-1}M)\right]\nonumber \\
 & \leq B+\sum_{i=d+1}^{n}\left[H(Y_{i-d}|Y^{i-d-1}X^{i-1})+H(X_{i}|Y^{i-d}M)\right],\label{eq:app13}
\end{align}
 where $B\triangleq\sum_{i=1}^{d}H(X_{i}|X^{i-1}M)+\sum_{i=n-d+1}^{n}H(Y_{i}|Y^{i-1}X^{n}M)$
is a finite constant that does not increase with $n.$ The inequality
in (\ref{eq:app13}) follows from conditioning reduces entropy. Note
also that we have the inequality $B\leq A$ by conditioning reduces
entropy.

By definition, a code $(d,n,R,0,\epsilon,d_{\max})$ must satisfy
(cf. (\ref{dist constraints-1}))
\begin{equation}
\epsilon\geq\frac{1}{n}\sum_{i=1}^{n}P_{e,i}\geq\frac{1}{n}\sum_{i=d+1}^{n}P_{e,i},\label{eq:app13b}
\end{equation}
 where we have defined $P_{e,i}\triangleq\Pr[X_{i}\neq Z_{1i}]$.
It follows that
\begin{align}
\sum_{i=d+1}^{n}H(X_{i}|Y^{i-d}M) & \leq\sum_{i=d+1}^{n}H(X_{i}|Z_{1i})\label{eq:app13a}\\
 & \leq\sum_{i=d+1}^{n}H_{b}(P_{e,i})+P_{e,i}\log|\mathcal{X}|\label{eq:app14}\\
 & \leq nH_{b}(\epsilon)+n\epsilon\log|\mathcal{X}|\label{eq:app15}\\
 & =\delta(\epsilon).\label{eq:app16}
\end{align}
The first inequality (\ref{eq:app13a}) follows from the fact that
$Z_{1i}$ is a function of $Y^{i-d}\mbox{ and }M$ by (\ref{eq:decoder})
and by conditioning reduces entropy; the second inequality (\ref{eq:app14})
follows from Fano's inequality and the third from (\ref{eq:app13b}).

Finally, from (\ref{eq:app11}),(\ref{eq:app12}),(\ref{eq:app13}),(\ref{eq:app16})
we obtain
\begin{align*}
nR & \geq A+\sum_{i=d+1}^{n}\left[H(Y_{i-d}|Y^{i-d-1}X^{i-1})+H(X_{i}|Y_{i-d}X_{i-d+1}^{i-1})\right]\\
 & -B-\sum_{i=d+1}^{n}\left[H(Y_{i-d}|Y^{i-d-1}X^{i-1})+n\delta(\epsilon)\right]\\
 & =A-B+\sum_{i=d+1}^{n}H(X_{i}|Y_{i-d}X_{i-d+1}^{i-1})+n\delta(\epsilon),
\end{align*}
 which concludes the proof.\hfill \QED

\section{Proof of Converse for Proposition \ref{pro:2} and Proposition \ref{pro:For-any-delay}\label{sub:Proof-of-Converse-2}}

We prove the converse for Proposition \ref{pro:For-any-delay}, since
Proposition \ref{pro:2} follows as a special case. We focus on $d=1$,
since the proof for $d=0$ can be obtained in a similar fashion. To
this end, fix a code $(1,n,R,\Delta R,D_{1}+\epsilon,D_{2}+\epsilon)$
as defined in Sec. \ref{sec:System-Model}. Using the definition of
encoder (\ref{eq:encoder}) and decoder (\ref{eq:decoder}) we have
\begin{align}
nR & \geq H(M)=I(M;X^{n}Y^{n})\nonumber \\
 & =\sum_{i=1}^{n}I(M;Y^{n})+I(M;X^{n}|Y^{n})\nonumber \\
 & =\sum_{i=1}^{n}I(M;Y_{i}|Y^{i-1})+I(M;X_{i}|Y^{n}X^{i-1})\nonumber \\
 & =\sum_{i=1}^{n}H(Y_{i}|Y_{i-1})-H(Y_{i}|Y^{i-1}M)+H(X_{i}|Y^{n}X^{i-1})-H(X_{i}|Y^{n}X^{i-1}M)\nonumber \\
 & =\sum_{i=1}^{n}H(Y_{i}|Y_{i-1})-H(Y_{i}|Z_{1i}Y^{i-1}M)+H(X_{i}|Y_{i})-H(X_{i}|Z_{1i}U_{i}Y_{i}Y_{i-1})\nonumber \\
 & \geq\sum_{i=1}^{n}H(Y_{i}|Y_{i-1})-H(Y_{i}|Z_{1i}Y_{i-1})+H(X_{i}|Y_{i}Y_{i-1})-H(X_{i}|Z_{1i}U_{i}Y_{i}Y_{i-1})\label{eq:app31}\\
 & =\sum_{i=1}^{n}I(Y_{i};Z_{1i}|Y_{i-1})+I(X_{i};Z_{1i}U_{i}|Y_{i}Y_{i-1}).\label{eq:app32}
\end{align}
 where we have defined $U_{i}\triangleq[Y_{1}^{i-2}Y_{i+1}^{n}X^{i-1}M]$.
All equalities above follow from standard properties of the entropy
and mutual information, while the inequality (\ref{eq:app31}) follows
by conditioning reduces entropy. Following the similar steps, we obtain
\begin{align}
n(R+\Delta R) & \geq H(M)+H(M_{\Delta})\geq H(MM_{\Delta})=I(MM_{\Delta};X^{n}Y^{n})\nonumber \\
 & =\sum_{i=1}^{n}I(MM_{\Delta};Y^{n})+I(M;X^{n}|Y^{n})\nonumber \\
 & =\sum_{i=1}^{n}H(Y_{i}|Y_{i-1})-H(Y_{i}|Y^{i-1}MM_{\Delta})+H(X_{i}|Y^{n}X^{i-1})-H(X_{i}|Y^{n}X^{i-1}MM_{\Delta})\nonumber \\
 & =\sum_{i=1}^{n}H(Y_{i}|Y_{i-1})-H(Y_{i}|Z_{1i}Y^{i-1}MM_{\Delta})+H(X_{i}|Y_{i})-H(X_{i}|Z_{1i}Z_{2i}U_{i}Y_{i}Y_{i-1}M_{\Delta})\nonumber \\
 & \geq\sum_{i=1}^{n}H(Y_{i}|Y_{i-1})-H(Y_{i}|Z_{1i}Y_{i-1})+H(X_{i}|Y_{i}Y_{i-1})-H(X_{i}|Z_{1i}Z_{2i}U_{i}Y_{i}Y_{i-1})\nonumber \\
 & =\sum_{i=1}^{n}I(Y_{i};Z_{1i}|Y_{i-1})+I(X_{i};Z_{1i}Z_{2i}U_{i}|Y_{i}Y_{i-1}).\label{eq:app4}
\end{align}
 The proof is concluded by introducing a time-sharing variable $T$
uniformly distributed in $[1,n]$ and defining random variables $X\triangleq X_{T},$
$Y\triangleq Y_{T}$, $Y_{1}\triangleq Y_{T-1}$, $Z_{1}=Z_{1T}$
and $Z_{2}=Z_{2T}$, and by leveraging the convexity of the mutual
informations in (\ref{eq:app32}) and (\ref{eq:app4}) with respect
to the distribution $p(z_{1i},z_{2i},u_{i}|x_{i},y_{i},y_{i-1})$.\hfill \QED

\section{Proof of (\ref{eq:ex1})-(\ref{eq:ex2})\label{sec:Proof-of-()-()}}

Here we prove that (\ref{eq:ex1})-(\ref{eq:ex2}) equals (\ref{eq:simpler})
for the binary hidden Markov model of Sec. \ref{sub:Binary-Hidden-Markov}.
First, for $D_{1}\geq\min\{\varepsilon^{(d)}*q,\mbox{ }1-\varepsilon^{(d)}*q\}=\varepsilon^{(d)}*q$,
we can simply set $Z_{1}=Y_{d}$ to obtain $I(X;Z_{1}|Y_{d})=0$ and
$E[X\bigoplus Z_{1}]\leq D_{1}$, which, from (\ref{eq:ex1}) and
the non-negativity of mutual information, leads to $R_{d}^{(a)}(D_{1})=0$.
Similarly, for $D_{1}\geq\min\{\varepsilon^{(d)}*q,\mbox{ }1-\varepsilon^{(d)}*q\}=1-\varepsilon^{(d)}*q$,
we can set $Z_{1}=1\oplus Y_{d}$ to prove that $R_{d}^{(a)}(D_{1})=0$.
For the remaining distortion levels $D_{1}\leq\min\{\varepsilon^{(d)}*q,\mbox{ }1-\varepsilon^{(d)}*q\}$,
under the constraint that $E[X\bigoplus Z_{1}]\leq D_{1}$, we have
the following inequalities
\begin{align}
I(X;Z_{1}|Y_{d}) & =H(X|Y_{d})-H(X|Y_{d}Z_{1})\\
 & =H_{b}(\varepsilon^{(d)}*q)-H_{b}(X\oplus Z_{1}|Y_{d}Z_{1})\\
 & \geq H_{b}(\varepsilon^{(d)}*q)-H_{b}(X\oplus Z_{1})\\
 & \geq H_{b}(\varepsilon^{(d)}*q)-H_{b}(D_{1}),
\end{align}
 where the third line follows by conditioning decreases entropy and
the last line from the fact that $H(x)$ is increasing in $x$ for
$x\leq1/2$. This lower bound can be achieved in (\ref{eq:simpler})
by choosing the test channel $p(z_{1}|x,y_{d})$ so that $X$ can
be written as
\begin{equation}
X=Y_{d}\oplus S\oplus Z_{1},
\end{equation}
 where $S$ is binary with $p_{S}(1)=D_{1}$ and independent of $Z_{1}$
and $Y_{d}$, and $Z_{1}$ is also independent of $Y_{d}$. To obtain
$p_{z_{1}}(1)$, we need to impose that the joint distribution $p(x,y_{d})$
is preserved by the given choice of $p(z_{1}|x,y_{d})$. To this end,
note that the joint distribution $p(x,y_{d})$ is such that we can
write $X=Y_{d}\oplus Q$, where $Q$ is binary and independent of
$Y_{d}$, with $p_{Q}(1)=\varepsilon^{(d)}*q$. Therefore, preservation
of $p(x,y_{d})$ is guaranteed if the equality $\Pr[S\oplus Z_{1}=1]=p_{z_{1}}(1)*D_{1}=\varepsilon^{(d)}*q$
holds. This leads to
\begin{equation}
p_{z_{1}}(1)=\frac{\varepsilon^{(d)}*q-D_{1}}{1-2D_{1}}.
\end{equation}
 We remark that $0\leq p_{z_{1}}(1)\leq1$, due to the inequality
(\ref{eq:ex2}) on the distortion $D_{1}$. This concludes the proof.\hfill \QED

\section{Proof of (\ref{eq:gauss r})\label{sec:Proof-of-()}}

Here we prove that (\ref{eq:gauss r}) equals (\ref{eq:simpler})
for the hidden Gauss-Markov model of Sec. \ref{sub:Gaussian-Hidden-Markov}.
This follows by using analogous arguments as done above for the binary
hidden Markov model. The only non-trivial adaptation of the proof
given above is the choice of the test channel for the case where $D_{1}\leq1-\rho^{2d}+\sigma_{N}^{2}$.
This must be selected so that $X$ can be written as
\begin{equation}
X=\rho^{d}Y_{d}+S+Z_{1},
\end{equation}
 where $S$ is zero-mean Gaussian with $E[S^{2}]=D_{1}$ and independent
of $Z_{1}$ and $Y_{d}$, and $Z_{1}$ is also zero-mean Gaussian
and independent of $Y_{d}$. To obtain $E[Z_{1}^{2}]$, we need to
impose that the joint distribution of $X$ and $Y_{d}$ is preserved
by the given choice of the test channel. To this end, note that the
joint distribution of $X$ and $Y_{d}$ is such that we can write
$X=\rho^{d}Y_{d}+Q+N$, where $Q$ is zero-mean Gaussian and independent
of $Y_{d}$ and $N$, with $E[Q^{2}]=1-\rho^{2d}$. Therefore, preservation
of the joint distribution of $X$ and $Y_{d}$ is guaranteed if the
equality $E[Z_{1}^{2}]+D_{1}=1-\rho^{2d}+\sigma_{N}^{2}$ holds. This
leads to
\begin{equation}
E[Z_{1}^{2}]=1-\rho^{2d}+\sigma_{N}^{2}-D_{1}.
\end{equation}
 We remark that $0\leq E[Z_{1}^{2}]\leq1$, due to the assumed inequality
on the distortion $D_{1}$.\hfill \QED

\end{document}